\documentclass{article}

\usepackage{epstopdf}
\usepackage{algorithmic}
\usepackage[cmex10]{amsmath}
\usepackage{url}
\usepackage{amssymb}
\usepackage{array}
\newcolumntype{C}[1]{>{\centering\let\newline\\\arraybackslash\hspace{0pt}}m{#1}}
\usepackage{graphicx}
\usepackage{grffile}
\usepackage[ruled,vlined]{algorithm2e}
\usepackage{subfigure}
\usepackage{threeparttable}
\usepackage{multirow}
\usepackage{mathrsfs}
\usepackage{lipsum}
\usepackage{color}
\usepackage{blindtext}
\usepackage{setspace}
\usepackage[T1]{fontenc}
\newtheorem{theorem}{Theorem}
\usepackage{float}
\usepackage{CJK}
 \usepackage[table,xcdraw]{xcolor}
 \renewcommand{\arraystretch}{1.4}
 \usepackage{booktabs}
\usepackage{amsmath}
\usepackage{cases}
\usepackage{color}
\usepackage{adjustbox}
\usepackage{rotating}

\usepackage{authblk}
\usepackage{bm}

\usepackage{cite}
\usepackage[numbers,sort&compress]{natbib} 
 
\usepackage{float}
\usepackage{verbatim}
\usepackage{float}

\usepackage[normalem]{ulem}

\usepackage{xcolor}
\usepackage{color, soul}

\newtheorem{definition}{Definition}
\newcommand{\nop}[1]{}

\begin{document}
\begin{sloppy}
%
\title{Applications of Differential Privacy in Social Network Analysis: A Survey}
\author[1]{Honglu~Jiang}
\author[2]{Jian~Pei}
\author[3]{Dongxiao~Yu \thanks{Corresponding author}}
\author[4]{Jiguo~Yu \thanks{Corresponding author}}
\author[5]{Bei~Gong}
\author[1]{Xiuzhen~Cheng}

\affil[1]{Department of Computer Science, The George Washington University, \{hljiang0720,cheng\}@gwu.edu} 
\affil[2]{School of Computing Science, Simon Fraser University, jpei@cs.sfu.ca}
\affil[3]{School of Computer Science \& Technology, Shandong University, dxyu@sdu.edu.cn}
\affil[4]{School of Computer Science \& Technology, Qilu University of Technology (Shandong Academy of Sciences), jiguoyu@sina.com}
\affil[5]{School of Information Technology, Beijing University of Technology, gongbei@bjut.edu.cn} 
\renewcommand*{\Affilfont}{\small\it}


\maketitle

\begin{abstract}

Differential privacy is effective in sharing information and preserving privacy with a strong guarantee. As social network analysis has been extensively adopted in many applications, it opens a new arena for the application of differential privacy.  
In this article, we provide a comprehensive survey connecting the basic principles of differential privacy and applications in social network analysis. We present a concise review of the foundations of differential privacy and the major variants and discuss how differential privacy is applied to social network analysis, including privacy attacks in social networks, types of differential privacy in social network analysis, and a series of popular tasks, such as degree distribution analysis, subgraph counting and edge weights.  We also discuss a series of challenges for future studies.

\nop{
This article surveys the foundations of differential privacy and t differential privacy and local differential privacy, and an overview on varieties of differentially private techniques and their applications in social networks. More specifically, we first detail the related concepts of differential privacy, including privacy budget, sensitivities, and noise mechanisms, with intuitive examples and explanations, to provide insights on the developments of various differential privacy techniques. Then we present the differential privacy standards for network structures and the diverse differentially private analysis techniques for social networks. We analyze the techniques targeting differentially private social network data analysis problems such as degree distribution, subgraph counting, and edge weight, following the four graph privacy standards of node privacy, edge privacy, out-link privacy, and partition privacy. Finally, according to the state-of-the-art research in differential privacy and their applications in social networks, we put forward four challenging problems to foster future studies on novel differential privacy techniques in social network analysis. 
}
\end{abstract}



\section{Introduction}\label{sec:introduction}

%
%
%
%
As a reflection of real social life, social networking shares and exchange a lot of private and sensitive information~\cite{NE}.  For example, in many online social networking sites, a user is required to provide personal information such as name, gender, birthdate, education level, marital status, personal photo, or even cell phone number. Besides, user-generated contents such as texts, pictures, videos, geographical locations published by users are also retained in the databases~\cite{BBS}. Such data is often shared with some third parties for services such as data analysis, targeted advertising, recommendations and evaluations on apps. If the personal private information is leaked or abused, the involved individuals may become subjects of intrusion attacks, such as spam mails, junk messages and telephone harassments. In some extreme cases, damages to personal reputation, properties, or even physical injuries may be caused due to illegal data disclosures~\cite{AAW}. 

The problem of data privacy protection was first put forward by Dalenius in the late 1970s~\cite{DT}, who pointed out that the purpose of protecting private information in a database is to prevent any user, including legitimate ones and potential attackers, from obtaining accurate information about arbitrary individuals when accessing the database. Following this principle, many privacy preservation models with strong operability were proposed, including $k$-anonymity~\cite{SLA}, $l$-diversity~\cite{MKV}, $t$-closeness~\cite{LLV} and $(\alpha,k)$-anonymity~\cite{WLF}. However, each of those models provides protection against only a specific type of attacks and cannot defend against newly developed ones. A fundamental cause of this deficiency lies in that the security of a privacy preservation model relies on an assumption of some specific background knowledge of an attacker. Nevertheless, it is almost impossible to enumerate all possible types of background knowledge that an attacker may have. Therefore, a privacy preserving model independent of background knowledge is highly desirable. \nop{Moreover, those privacy preservation models, such as $k$-anonymity and $l$-diversity, cannot rigorously prove their level of privacy preservation. When the model parameters are changed, it is hard to quantitatively analyze their levels of privacy preservation, weakening the reliability of the processed query results~\cite{XZW}.}

\begin{sidewaystable}
\caption{Summary on Previous Survey Articles} 
\scalebox{0.85}{
\begin{tabular}{|c|c|l|l|}
\hline
\multicolumn{1}{|c|}{\textbf{Topic}}  & \textbf{Ref}  & \textbf{Focuses}                                                                                                                                                                                                          & \textbf{Major angles} \\
\hline 
\hline                           
\multirow{6}{*}{\begin{tabular}[c]{@{}c@{}}\textbf{Differential}\\ \textbf{Privacy} \\ \textbf{Techniques}\end{tabular}} &~\cite{dwork2008}    & \begin{tabular}[l]{@{}l@{}}Basic techniques to achieve differential privacy and applications \\ in statistical datasets.\end{tabular}                                                                                    & \begin{tabular}[c]{@{}l@{}}Learning theory,\\ Statistical datasets\end{tabular}                    \\
\cline{2-4}
  & \begin{tabular}[c]{@{}l@{}}  ~\cite{zhu2017} \end{tabular}             & \begin{tabular}[c]{@{}l@{}}Differential privacy theory and application on two aspects of statistical\\ datasets, privacy preserving data release and privacy preserving data mining.\end{tabular} & \begin{tabular}[c]{@{}l@{}}Statistical data publishing and mining,\\ Applications of differential privacy\end{tabular}     \\

  \cline{2-4}
&~\cite{hassan}            &\begin{tabular}[c]{@{}l@{}} A comprehensive survey on differential privacy techniques for \\cyber-physical systems.\end{tabular}.   & \begin{tabular}[c]{@{}l@{}}Application and implementation\\ in cyber-physical systems\end{tabular} \\
\hline
\multirow{6}{*}{\begin{tabular}[c]{@{}c@{}}\textbf{Privacy of}\\ \textbf{Social} \\ \textbf{Networks}\end{tabular}}            & \begin{tabular}[c]{@{}l@{}}\cite{zhou2008, zheleva2011} \\~\cite{ssp2012, AJH}\end{tabular}     & \begin{tabular}[c]{@{}l@{}}Surveys on privacy risks and anonymization techniques for privacy \\ preserving publishing of social network data.\end{tabular}                                                                             & \begin{tabular}[c]{@{}l@{}}Privacy risks,\\ Anonymization in social networks\end{tabular}          \\
\cline{2-4}
&~\cite{ji2016}                & \begin{tabular}[c]{@{}l@{}}A survey on graph data anonymization, de-anonymization attacks \\and de-anonymizability quantification.\end{tabular}                                                                          & \begin{tabular}[c]{@{}l@{}}Anonymization,\\ De-anonymization\end{tabular}                          \\
  \cline{2-4}
  &~\cite{beigi2020}         & \begin{tabular}[c]{@{}l@{}}A review on the privacy risks that exist in different aspects of\\ social media data such as attribute and identity disclosure attacks.\end{tabular}                                                                                   & \begin{tabular}[c]{@{}l@{}}Privacy risks,\\ Social media\end{tabular}\\                         
\hline
\end{tabular}}
\label{tab:survey}
\end{sidewaystable}

Dwork developed \emph{differential privacy}~\cite{DCD} to provide a strong privacy guarantee and protect against the privacy disclosure of statistical databases. Under differential privacy, query results of a dataset are insensitive to the change of a single record. That is, whether a single record exists in the dataset has little effect on the output distribution of the analytical results. An attacker cannot obtain accurate individual information by observing the results because the risk of privacy disclosure generated by adding or deleting a single record is kept within a user-specified, acceptable range. Differential privacy 
assumes that an attacker can obtain all information in a dataset except for the target record, which can be regarded as the maximum background knowledge that an attacker can have. 
\nop{Under this assumption, differential privacy does not need to take into account the background knowledge possessed by an attacker as it cannot be more abundant than the assumed maximum one.} 
It rests on a sound mathematical foundation
under certain assumptions as well as quantitative evaluations. Differential privacy is a standard for quantifying privacy risks rather than a single tool and has been widely used in statistical estimations, data publishing, data mining and machine learning. There exist many methods and implementations to achieve differentially private data analysis.

Differential privacy mainly aims at statistical problems in databases at first. Because of its unique strengths, differential privacy has been applied to social network data analysis. A number of suitable adaptations of differentially private social network analysis techniques have been developed~\cite{HLM,dll,knrs,NRS,SZW,BBD,ZTY}. 
Social networks post a series of challenges for privacy preserving analysis.
Social networks can be modeled as graphs and become very complicated at large scale. They often have high data correlations since social relationships among users are not independent. 
As demonstrated by Liu~\emph{et~al.}~\cite{LCM}, the dependence among tuples in statistical databases may seriously weaken the privacy guarantee that current differential privacy mechanisms provide. This challenge obviously also holds for social networks. 

There exist at least three fundamental challenges that need to be tackled in order to apply differential privacy to social network analysis.  First, we have to adapt differential privacy from tabular data to network data.  Second, we have to address the issue of high sensitivity in complex and correlated social network data.  Last, we have to explore the tradeoff between data utility and privacy guarantee as too much noise may make query results useless.
%


Understanding differential privacy and applications in social network analysis comprehensively is far from trivial. 
There exist multiple relevant surveys on differential privacy~\cite{ dwork2008, zhu2017, hassan} and privacy preservation in social network analysis~\cite{zhou2008,ssp2012,zheleva2011,AJH,ji2016,beigi2020}, whose topics, focuses and major angles are summarized in Table~\ref{tab:survey}. Those existing surveys focus on either differential privacy in tabular statistical databases or privacy preservation on social network analysis, but none of them closely connects the two in a comprehensive and integrated manner.  This motivates our endeavor in this article, whose major objective is to provide detailed interpretations and intuitive illustrations on differential privacy foundations, especially for noise calibration to global sensitivity and smooth sensitivity, and then extend them to the state-of-the-art differentially private social network analysis techniques addressing the three major challenges mentioned above.

Conducting research on differential privacy in social networks needs real social network data. The Stanford Network Analysis Platform (SNAP)\footnote{\url{http://snap.stanford.edu/data/}} provides an extensive repository~\cite{snap}. It includes some popular online social networks, communication networks, citation networks, web and blog datasets, and several other large network datasets.

The rest of the article is organized as follows.
In Section~\ref{sec:pre}, we review the basic concepts of differential privacy with detailed interpretations and examples. More specifically, we define and explain the differential privacy model, describe its noise mechanisms calibrated to global sensitivity and smooth bounds of local Sensitivity, and present the composition properties. For better elaboration, we exemplify popular functions such as \textsf{count} and \textsf{median}, and detail the corresponding differentially private noise mechanisms. 

In Section \ref{sec:var}, we discuss two most popular variants of differential privacy. \emph{Dependent differential privacy} is proposed to handle queries involving correlated database tuples. \emph{Local differential privacy} is a well-developed extension to centralized differential privacy.

In Section~\ref{sec:dpg}, to effectively demonstrate how to adapt differential privacy from tabular data to social network data, we first summarize the popular privacy attacks in social networks, and then introduce the four types of network privacy, namely \emph{node privacy}, \emph{edge privacy}, \emph{out-link privacy}, and \emph{partition privacy}. We illustrate the definitions of these graph differential privacy types and analyze their applicability and complexity.

In Section~\ref{sec:snat}, we provide an overview on differentially private algorithms for \emph{degree distribution}, \emph{subgraph counting}, and \emph{edge weight}, the three most widely-used graph analysis techniques under the types of social network privacy mentioned above in Section~\ref{sec:dpg}. Our analysis demonstrates that most of them cannot obtain good utility due to large network size, complex graph structures and strong graph attribute correlations. 
\nop{For certain graph queries, such as $k$-triangle counting~\cite{KRS}, query answering itself is a hard problem, let alone the corresponding differentially private algorithms that require the computations of sensitivity values. Moreover, although the high sensitivity problem is not severe when considering local differential privacy, it is a great challenge for data collectors to reconstruct a graph structure with high utility based on the disturbed or local graph data, and the generated graph may not retain the important characteristics of the original one.}

In Section~\ref{sec:saf}, we conclude this article and describe a few open research challenges.

\section{Differential Privacy}\label{sec:pre}

In this section, we review the core concepts in differential privacy.
For better elaboration of noise mechanisms, we exemplify popular query functions, such as \textsf{count} and \textsf{median}, to illustrate the corresponding noise mechanisms calibrated to \emph{global sensitivity} and \emph{smooth sensitivity}. 

\subsection{Intuition}

An individual's information may be inferred even without explicitly querying for the specific details. For example, consider the data in Table~\ref{tab:1}, which is about whether a person suffers from a disease.  Suppose the database provides a query interface $Q_{i}(D)$, which returns the sum of the second column, `Disease or Not', of the first $i$ rows. The query returns an aggregate and does not query any specific person.

\begin{table}[t]
\small
\renewcommand{\arraystretch}{1.0}
\arrayrulewidth=1pt   
\tabcolsep=14pt   
\setlength{\belowcaptionskip}{10pt}%
\caption{An Example Database}
\centering
\begin{tabular}{|c|c|}
\rowcolor[gray]{.8}
\hline
 \textbf{Name}  & \textbf{Disease or Not}\\
\hline Ross    & 1\\
\hline Monica    & 1\\
\hline Bob      & 0\\
\hline Joey   & 0\\           
\hline Alice       &1\\
\hline
\end{tabular}
 \label{tab:1}
\end{table}

Suppose an attacker wants to infer whether or not Alice has the disease with the background knowledge that the record about Alice is the last one in the database. The attacker can issue two queries $Q_{5}(D)$ and $Q_{4}(D)$, and compute the difference of the results, $Q_{5}(D)-Q_{4}(D)$. Alice has the disease if the outcome is $1$ and she does not have the disease otherwise. This simple example shows how personal information may be disclosed even when it is not explicitly queried. It is not safe to release the exact query answers even when data is not published. 

The intuition of differential privacy is to inject a controlled level of statistical noise into query results to hide the consequence of adding or removing an arbitrary individual from a dataset. That is, when querying two almost identical datasets (differing by only one record, for example), the results are differentially privatized so that an attacker cannot glean any new knowledge about an individual with a high probability, i.e., whether or not a given individual is present in the dataset cannot be guessed with useful confidence. In the example shown in Table~\ref{tab:1}, to protect Alice's privacy, we can inject noises into answers to $Q_{5}(D)$ and $Q_{4}(D)$ so that $Q_{5}(D)-Q_{4}(D)$ and Alice's value on the column `Disease or Not' are independent with high probability.

\subsection{Definition of Differential Privacy}

Let $f$ be a query function to be evaluated on a dataset $D$. We want to have an algorithm $A$ running on the dataset $D$ and returning $A(D)$ such that $A(D)$ should be $f(D)$ with a controlled amount of random noise added. The goal of differential privacy is to make $A(D)$ close to $f(D)$ as much as possible to ensure data utility, and at the same time $A(D)$ preserves the privacy of the entities in the dataset.

Differential privacy mainly addresses adversarial attacks that queries datasets differing by only a small number of entries. There are two flavors of differential privacy, namely \emph{unbounded} and \emph{bounded}, which are distinguished by the definition of \emph{neighboring datasets}~\cite{DKA}. For two datasets $D$ and $D'$, if $D'$ can be obtained by adding or removing a tuple from $D$, it is called unbounded. If $D'$ can be obtained by changing the value of a tuple from $D$, then it is called bounded. That is, bounded neighboring datasets have the same size while the sizes of two unbounded neighboring datasets differ by $1$. 
There exist slight differences in presenting the query results for unbounded and bounded neighboring datasets, but
 the ideas of designing and analyzing the differential privacy mechanisms are the same. Therefore in this article, we employ both types of neighboring datasets to illustrate the introduced differential privacy mechanisms.

\begin{definition}[Differential privacy~\cite{DMN}] \label{def:DP}
A randomized algorithm $A$ is \emph{$\epsilon$-differentially private} if for any two neighboring datasets $D$ and $D'$, and any subset $S$ of possible outputs of $A$,
\begin{equation}
Pr[A(D)\in S]\leq e^{\epsilon}Pr[A(D')\in S],\nonumber
\end{equation}
where $\epsilon \geq 0$ is a parameter called \emph{privacy budget}.
\end{definition}


Privacy budget $\epsilon$ in Definition~\ref{def:DP} is often a small positive real number and reflects the level of privacy preservation that algorithm $A$ can provide. For example, if $\epsilon=0.01$, $e^{0.01}\approx 1.01$; and $0.01$-differential privacy ensures that the distributions of $A(D)$ and $A(D')$ are very similar and almost indistinguishable.
The smaller the value of $\epsilon$, the higher the level of privacy preservation. A smaller $\epsilon$ provides greater privacy preservation at the cost of lower data accuracy since more noise has to be added. When $\epsilon=0$, the level of privacy preservation reaches the maximum, that is, ``perfect'' protection. In this case, the algorithm outputs two results with indistinguishable distributions but the corresponding results do not reflect any useful information about the dataset. Therefore, the setting of $\epsilon$ should balance the tradeoff between privacy and data utility. In practical applications, $\epsilon$ usually takes very small values such as $0.01$, $0.1$, or $\ln 2$, $\ln 3$~\cite{dwork2008}. 
Computing $\epsilon$-differential privacy may be challenging in some scenarios. To facilitate approximation, a generalized notion of differential privacy is developed.

\begin{definition}[Approximate differential privacy~\cite{DCKK}] \label{def:ADP}
A randomized algorithm $A$ is $(\epsilon, \delta)$-differentially private if for any two neighboring datasets $D$ and $D'$, and any subset $S$ of possible outputs of $A$,
\begin{equation}
Pr[A(D)\in S]\leq e^{\epsilon}Pr[A(D')\in S]+\delta\nonumber
\end{equation}

\end{definition}

When $\delta>0$, $(\epsilon, \delta)$-differential privacy relaxes $\epsilon$-differential privacy by a small probability controlled by parameter $\delta$. In $\epsilon$-differential privacy, the ratio between the output probability distributions for neighboring datasets $D$ and $D'$ is strictly bounded by $e^{\epsilon}$; while in $(\epsilon, \delta)$-differential privacy, a freedom to breach the strict $\epsilon$-differential privacy for certain low probability events is offered. That is, in $(\epsilon, \delta)$-differential privacy, equation $Pr[A(D)\in S]\leq e^{\epsilon}Pr[A(D')\in S]$ holds with the probability at least $1-\delta$.

Typically, $\delta$ is set to far smaller than the inverse of any polynomial in the size $n$ of the database (i.e., $\delta \ll \frac{1}{p(n)}$)~\cite{dwork2014}. An equivalent formulation states that $\delta$ is cryptographically negligible when $\delta \leq n^{-\omega(1) }$~\cite{dwork2014}. Note that $\frac{1}{p(n)}$ can be described as an upper bound of $\delta$ since a value of $\delta$ in the order of $\frac{1}{p(n)}$ is dangerous for privacy leakage.


Differential privacy can be achieved by adding an appropriate amount of noise to query results, that is, $A(D)=f(D)+Z$, which is illustrated in Figure~\ref{fig:frame}. Adding too much noise may decrease data utility, while adding too little noise cannot provide sufficient privacy guarantee. Sensitivity, which represents the largest change to the query results caused by adding/deleting any record in the dataset, is the key parameter to determine the magnitude of the added noise. Accordingly, global sensitivity, local sensitivity, smoothing upper bound and smoothing sensitivity are defined under the differential privacy model.

 \begin{figure}[t]
 \centering
  \includegraphics[width=0.8\textwidth]{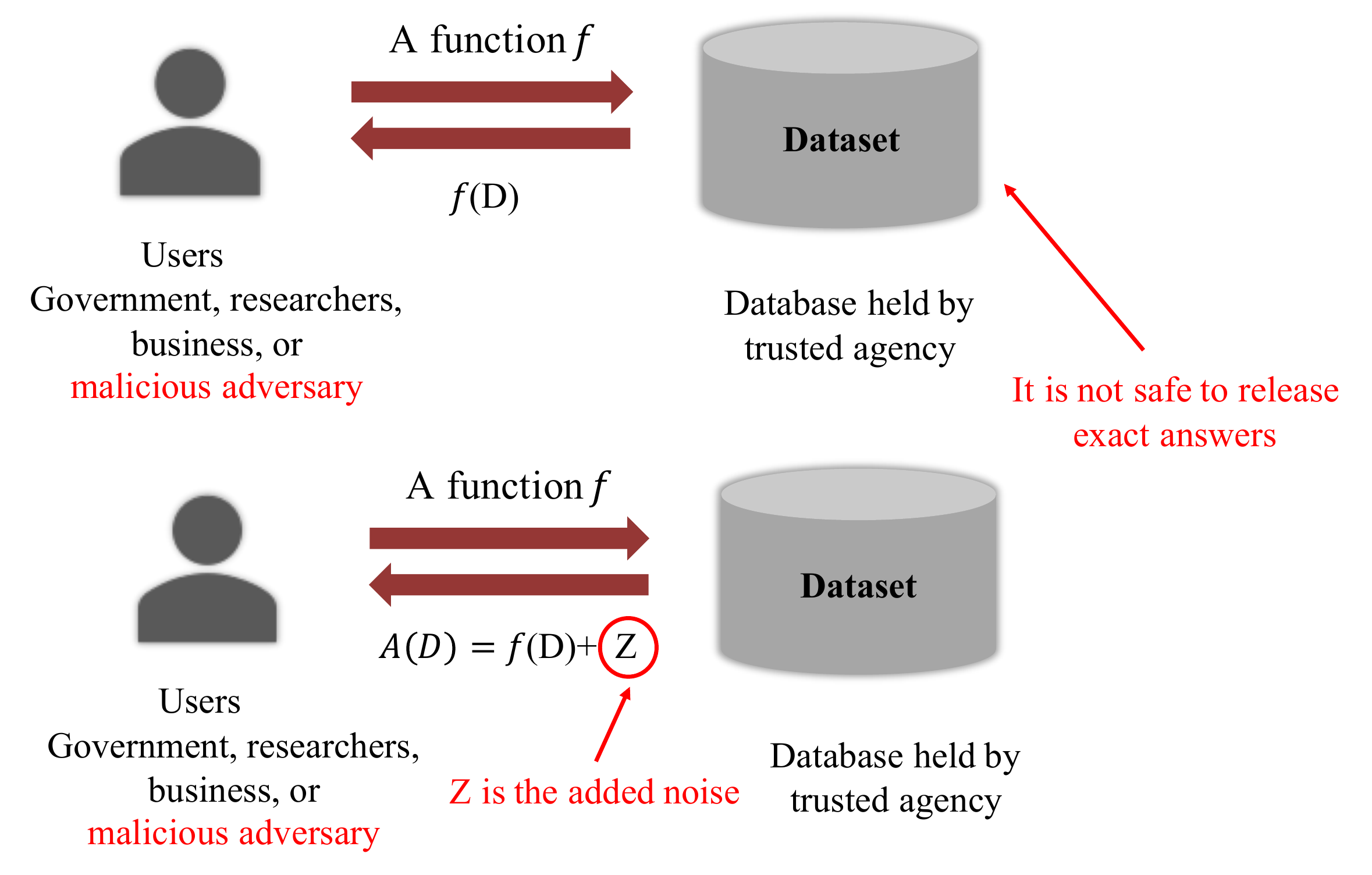}
  \caption{A framework of output perturbation in differential privacy.}
  \label{fig:frame}
\end{figure}

\subsection{Noise Calibration}

How to add noise to query results $f(D)$ and how much noise should be added are the key to the noise mechanism in differential privacy. In this subsection, we introduce two frameworks of differentially private noise mechanisms, namely, noise calibration to global sensitivity and that to smooth sensitivity.

\subsubsection{\textbf{Noise Calibration to Global Sensitivity}}\label{sec:global}

\begin{definition}[Global sensitivity~\cite{DCD}] \label{def:global:sensitivity}
For $f:D\rightarrow R^{d}$, the global sensitivity of $f$ for all pairs of neighboring datasets $D$ and $D'$ is 
$$GS_{f}=\max\limits_{D, D'}\|f(D)-f(D')\|_{1},$$
where $\| \cdot \|_{1}$ denotes the $L_{1}$ norm.
 \end{definition}
 
The global sensitivity measures the maximum change of query results when modifying one tuple. It is only related to the query function, and is independent from the dataset itself. 

For some functions such as \textsf{sum}, \textsf{count}, and \textsf{max}, the global sensitivity is easy to compute. For instance, the global sensitivity for counting is $1$ since only one tuple is changed for any two neighboring datasets, and that for the histogram query is $2$ as illustrated in Figure~\ref{fig:4}. 

\begin{figure}[t]
 \centering
  \includegraphics[width=0.8\textwidth]{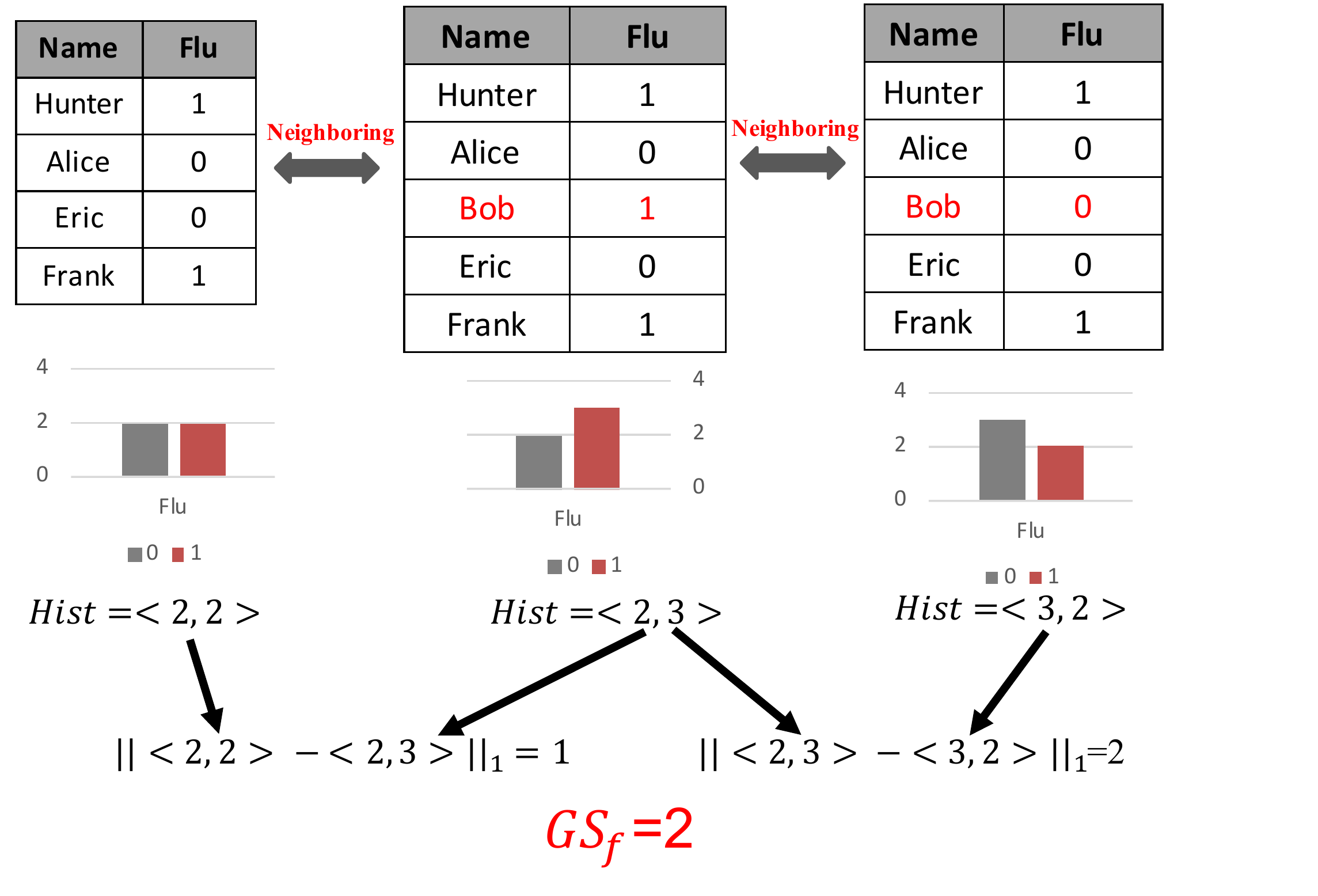}
  \caption{Global sensitivity of the histogram query.}
  \label{fig:4}
\end{figure}

For some other functions, such as calculating the maximum diameter of $k$-means clusters and counting subgraphs, the global sensitivity may be difficult to compute or unbounded. For example, the \textsf{median} function can have a high global sensitivity. Take $f(D)=median (x_{1}, x_{2}, \ldots, x_{n})$ as an example, of which $x_{i}$ is a real number in $[0,M]$. Assume that  $n$ is an odd number and that $x_{1}, x_{2}, \ldots, x_{n}$ are sorted. Thus, $f(D)=x_{m}$, where $m=\frac{n+1}2$. Consider the following extreme case,
\begin{eqnarray}
 D &:&  \{0,0,\ldots, x_{m}=0, x_{m+1}=M, \ldots, M\},\nonumber\\
 D' &:&  \{0,0,\ldots, x_{m}=M, \ldots, M\}.\nonumber
\end{eqnarray}
We have $f(D)=0$ and $f(D')=M$. Therefore, the global sensitivity for this function is $M$, which can be arbitrarily large.  As another example, the global sensitivity of the triangle counting query of a graph is unbounded, since the change of triangle counting depends on the graph size.

The noise injected to achieve differential privacy can be calibrated according to the global sensitivity of the query function, that is, the maximum amount of change to the query result when only one record is modified in the dataset. For a function with a small global sensitivity, only a small amount of noise needs to be added to cover up the impact on query results when one record is changed. However, when the global sensitivity is large, it is necessary to add a substantial amount of noise to the output to ensure the privacy guarantee, which leads to poor data utility. Two noise mechanisms, namely Laplace mechanism~\cite{DMN} and exponential mechanism~\cite{MTK}, were proposed for different problems.

\paragraph{Laplace Mechanism}

The Laplace distribution~\cite{laplace} (centered at $\mu$) with scale $b$ is the distribution with probability density function
$$h(z)=\frac{1}{2b}\exp(-\frac{|z-\mu|}{b}).$$
Denote by $Lap(b)$ the Laplace distribution (centered at $0$) with scale $b$. Dwork~\textit{et~al.}~\cite{DMN} proposed the Laplace mechanism, which states that for dataset $D$ and function $f: D\rightarrow R^{d}$ with global sensitivity $GS_{f}$, $A(D)=f(D)+Z$ is $\epsilon$-differentially private, where $Z\sim Lap(GS_{f}/\epsilon)$.

The Laplace mechanism is suitable for protecting numerical results. Figure~\ref{fig:6} shows an example of Laplace mechanism for the counting function. Since the global sensitivity of counting is $GS_{f}=1$, if we choose $\epsilon=0.1$, the Laplace mechanism outputs $3+Lap(10)$ for the specific $D$ in Figure~\ref{fig:6}.

\begin{figure}[t]
 \centering
  \includegraphics[width=0.8\textwidth]{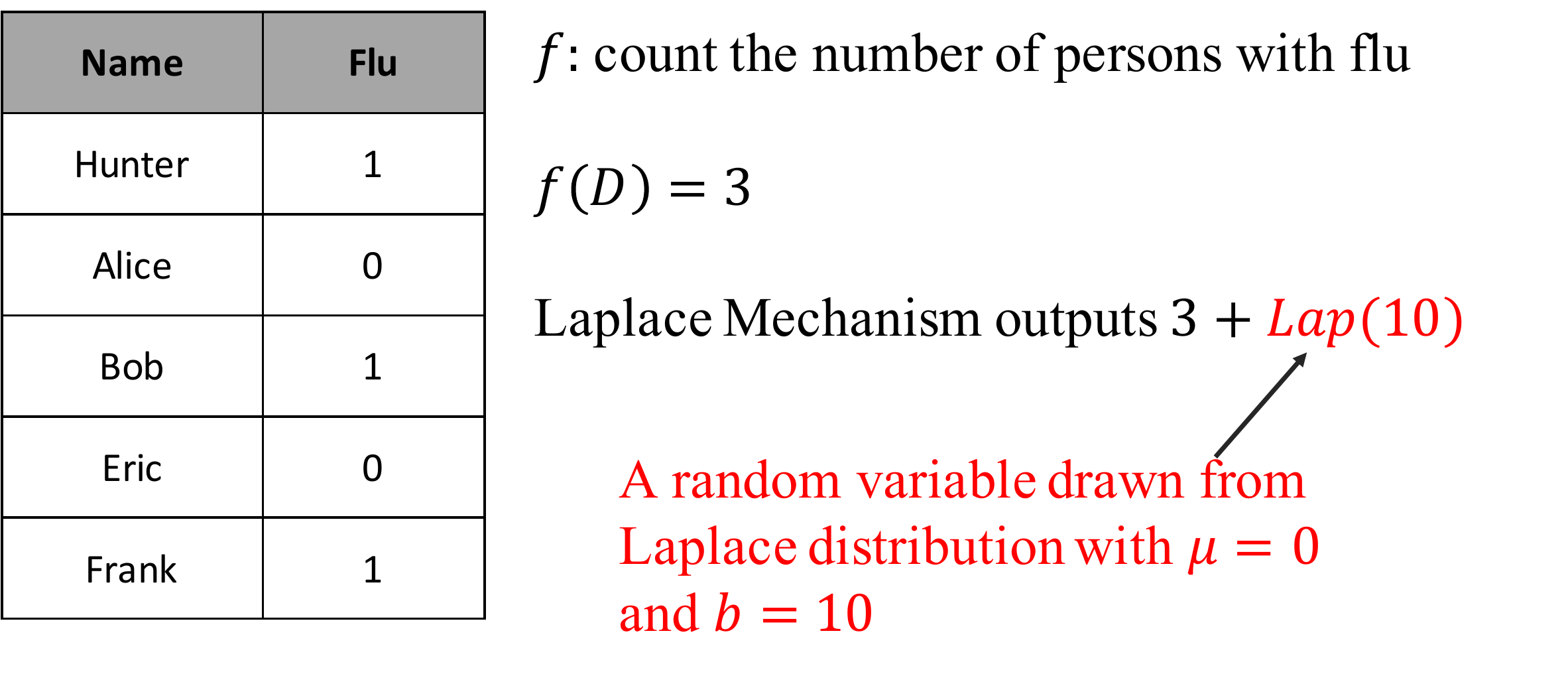}
  \caption{Laplace mechanism for the counting function.}
  \label{fig:6}
\end{figure}

\paragraph{Exponential Mechanism}

\begin{sidewaystable}
\begin{center}
\caption{An Example Exponential Mechanism} \label{tab:2}
\scalebox{0.9}{
\begin{tabular}{|m{1cm}<\centering|m{1.5cm}<\centering|m{6cm}<\centering|p{6cm}<\centering|p{4.5cm}<\centering|}
\hline
\multirow{2}{*}{\textbf{Item}} & \multirow{2}{*}{\textbf{\begin{tabular}[c]{@{}c@{}} $\bm{q(D,r)}$\\ ($\bm{\Delta q=1}$)\end{tabular}}} & \multicolumn{3}{c|}{\textbf{Probability}} \\\cline{3-5}
 & & \textbf{$\epsilon=0$} & \textbf{$\epsilon=0.1$}& \textbf{$\epsilon=1$} \\ \hline
$\mathcal{A}$ &10& \begin{tabular}[c]{@{}l@{}} $exp(\frac{0\times10}{2\times1})=1$, \\ $Pr\left[output=\mathcal{A}\right] = \frac{1}{1+1+1}=1/3$\end{tabular} & \begin{tabular}[c]{@{}l@{}}$exp(\frac{0.1\times10}{2\times1})=e^{0.5}$,\\ $Pr\left[output=\mathcal{A}\right]=\frac{e^{0.5}}{e^{0.5}+e^{1}+e^{1.5}}$\\$=0.186$\end{tabular}& \begin{tabular}[c]{@{}l@{}}$exp(\frac{1\times10}{2\times1})=e^{5}$,\\ $Pr\left[output=\mathcal{A}\right]=\frac{e^{5}}{e^{5}+e^{10}+e^{15}}$\\$=4.509\times 10^{-5}$ \end{tabular}\\\hline
$\mathcal{B}$ &20& \begin{tabular}[c]{@{}l@{}} $exp(\frac{0\times20}{2\times1})=1$, \\ $Pr\left[output=\mathcal{B}\right] = \frac{1}{1+1+1}=1/3$\end{tabular} & \begin{tabular}[c]{@{}l@{}}$exp(\frac{0.1\times20}{2\times1})=e^{1}$,\\ $Pr\left[output=\mathcal{B}\right]=\frac{e^{1}}{e^{0.5}+e^{1}+e^{1.5}}$\\$=0.307$\end{tabular} & \begin{tabular}[c]{@{}l@{}}$exp(\frac{1\times20}{2\times1})=e^{10}$,\\ $Pr\left[output=\mathcal{B}\right]=\frac{e^{10}}{e^{5}+e^{10}+e^{15}}$\\$=0.0686$\end{tabular} \\\hline
$\mathcal{C}$ &30& \begin{tabular}[c]{@{}l@{}} $exp(\frac{0\times30}{2\times1})=1$, \\ $Pr\left[output=\mathcal{C}\right] = \frac{1}{1+1+1}=1/3$\end{tabular} & \begin{tabular}[c]{@{}l@{}}$exp(\frac{0.1\times30}{2\times1})=e^{1.5}$,\\ $Pr\left[output=\mathcal{C}\right]=\frac{e^{1.5}}{e^{0.5}+e^{1}+e^{1.5}}$\\$=0.506$\end{tabular} & \begin{tabular}[c]{@{}l@{}}$exp(\frac{1\times30}{2\times1})=e^{15}$,\\ $Pr\left[output=\mathcal{C}\right]=\frac{e^{15}}{e^{5}+e^{10}+e^{15}}$\\$=0.993$\end{tabular}\\ \hline
\end{tabular}}
\end{center}
\end{sidewaystable}

In some situations, query results are categorical, such as finding the zip code of the highest average income. McSherry~\emph{et~al.}~\cite{MTK} developed the exponential mechanism for the situations where the ``best'' needs to be selected. Let $Range$ be the output domain of a query function and each value $r\in Range$ be an entity object. In the exponential mechanism, the \emph{utility function} of the output value $r$, denoted by $q(D,r)$, is employed to evaluate the quality of $r$.
Given a randomized algorithm $A$ with input dataset $D$ and output entity object $r\in Range$, let 
$\Delta q$ be the global sensitivity of function $q(D,r)$. McSherry~\emph{et~al.}~\cite{MTK} showed that, if an algorithm $A$ selects and outputs $r$ from $Range$ at a probability proportional to $\exp(\frac{\epsilon q(D,r)}{2\Delta q})$, then $A$ is $\epsilon$-differentially private.

Table~\ref{tab:2} presents an example of the exponential mechanism. There is a basket $\mathcal{D}$ with three kinds of fruits: apple ($\mathcal{A}$), banana ($\mathcal{B}$), and cherry ($\mathcal{C}$). Algorithm $A$ seeks to output the kind of fruits that has the largest amount. Let $q(D,\mathcal{A})=count(\mathcal{A})$, $q(D,\mathcal{B})=count(\mathcal{B})$, and $q(D,\mathcal{C})=count(\mathcal{C})$. Thus $\Delta q=1$, since adding or removing an apple, a banana or a cherry causes a change of the utility function value to be at most $1$. Based on the exponential mechanism, one can compute the probabilities of outputting $\mathcal{A}$, $\mathcal{B}$ and $\mathcal{C}$ with a given $\epsilon$, which are shown in Table~\ref{tab:2}.

The output probability of the item with a high utility function is amplified when $\epsilon$ is large, such as when $\epsilon=1$ in the table. As $\epsilon$ decreases, the utility differences of the items become more and more smoothed and the probabilities of the outputs tend to be equal. When $\epsilon=0$, the output probabilities for all items are equal.

\subsubsection{\textbf{Noise Calibration to Smooth Sensitivity}}

\noindent \quad When the global sensitivity is large, a substantial amount of noise has to be added to the output to achieve differential privacy, which may seriously impair data utility. To address this issue, Nissim~\textit{et~al.}~\cite{NRS} proposed the idea of local sensitivity.

\begin{definition}[Local sensitivity~\cite{NRS}] \label{def:local:sensitivity}
The local sensitivity of function $f: D\rightarrow R^{d}$ on $D$ is
$$LS_{f}(D)=\max\limits_{\text{neighboring data set }D' \text{ of }D}\|f(D)-f(D')\|_{1}$$
\end{definition}

Let us take the \textsf{median} function as an example, that is, $f(D)=median(x_{1}, x_{2}, \ldots, x_{n})$, where $n$ is odd. We have $f(D)=x_{m}$, where $m=\frac{n+1}2$, and $LS_{f}(D)=\max\{x_{m}-x_{m-1}, x_{m+1}-x_{m}\}$.

The local sensitivity is related to not only the query function $f$ but also the given dataset $D$. According to Definition~\ref{def:global:sensitivity}, $GS_{f}=\max\limits_{D}(LS_{f}(D))$. Since the magnitude of noise is proportional to sensitivity, the amount of noise added is much less with local sensitivity. Unfortunately, local sensitivity does not satisfy the requirement of differential privacy, because the noise magnitude itself may reveal the database information.  For example,  consider a database where the values are between $0$ and $M>0$, and two neighboring databases  $D(0,0,0,0,0,M,M)$ and $D'(0,0,0,0,M,M,M)$. Let $f$ be the median function. Then, $f(D)=0$ and $f(D')=0$, and the corresponding local sensitivities are $LS_{f}(D)=0$ and $LS_{f}(D')=M$. Correspondingly, if the noises are calibrated according to $0$ and $M$, respectively, to compute $A(D)$ and $A(D')$, then they are easy to be distinguished by an adversary.
An algorithm $A$ is not ($\epsilon, \delta$)-differentially private if local sensitivity is adopted. 

To bridge the gap, a \emph{smooth upper bound} of the local sensitivity is proposed to determine the magnitude of the added noise~\cite{NRS}.

\begin{definition}[Smooth bound and smooth sensitivity~\cite{NRS}] \label{def:smooth:bound}
 For a dataset $D$ and a query function $f$, a funtion $S: D\rightarrow R$ is a $\beta$-smooth upper bound of $LS_{f}(D)$ with $\beta>0$, if $\forall D, S(D)\geq LS_{f}(D)$ and $\forall D, D' \mbox{ with } d(D,D')=1,  S(D)\leq e^{\beta}S(D')$.

\label{def:smooth:sensitivity}
The $\beta$-smooth sensitivity of function $f$ with $\beta>0$ is 
$$S^{*}_{f,\beta}(D)=\max\limits_{D'}\{LS_{f}(D')\cdot e^{-\beta \cdot d(D,D')}\}.$$
\end{definition}

When $\beta=0$, $S(D)$ becomes the constant $GS_{f}$ to satisfy the requirements in Definition~\ref{def:smooth:bound}. Global sensitivity is a simple but possibly loose upper bound on $LS_{f}$. When $\beta>0$, global sensitivity is a conservative upper bound on $LS_{f}$. $LS_{f}$ may have multiple smooth bounds, and the smooth sensitivity is the smallest one that meets Definition~\ref{def:smooth:bound}.


Again, consider the \textsf{median} function as an example. We construct a function $A^{(k)}(D)$ that counts how much the sensitivity can change when up to $k$ entries are modified.
 $$A^{(k)}(D)=\max\limits_{D'\in\mathbb{D}:d(D,D')\leq k}LS_{f}(D')$$
 where $\mathbb{D}$ is the domain of all possible datasets.  Then, the smooth sensitivity can be expressed using $A^{(k)}(D)$ as
\begin{eqnarray}
S^{*}_{f,\beta}(D) &=&\max\limits_{k=0,\ldots,n}e^{-k\beta}(\max\limits_{D'\in\mathbb{D}:d(D,D')\leq k}LS_{f}(D')),\nonumber \\
& = &\max\limits_{k=0,\ldots,n}e^{-k\beta}A^{(k)}(D). \nonumber
\end{eqnarray}

To compute $A^{(k)}(D)$, we need to calculate the maximum of $LS_{f}(D')$ where $D'$ and $D$ differ by up to $k$ tuples. Recall that $D$ is sorted, $f(D)=x_{m}$, and $LS_{f}(D)=\max\{x_{m}-x_{m-1}, x_{m+1}-x_{m}\}$. Thus, we have

\begin{eqnarray}\label{eq:Ak}
A^{(k)}(D) &=&\max\limits_{D'\in\mathbb{D}:d(D,D')\leq k}LS_{f}(D') \nonumber \\ & = & \max\limits_{0 \leq t\leq k+1}\{x_{m+t}-x_{m+t-k-1}\}.\nonumber
\end{eqnarray}

Then, the smooth sensitivity of the median function can be calculated by

\begin{equation}
\begin{aligned}
S^{*}_{f_{med},\beta}(D)=\max\limits_{0\le k \le n}(e^{-k\beta}\cdot \max\limits_{0\le t \le k+1}(x_{m+t}-x_{m+t-k-1}))\nonumber
\end{aligned}
\end{equation}

In general, computing smooth sensitivities for functions such as the number of triangles in a graph is non-trivial and even NP-hard~\cite{KRS}. Therefore, a smooth upper bound is used to replace a smooth sensitivity when the latter is hard to compute. Next, we show how to employ a $\beta$-smooth sensitivity (or upper bound) to calibrate noise for $\epsilon$-differential privacy.

According to the framework of differential privacy presented in Section~\ref{sec:global}, $A(D)=f(D)+ Z$ is returned for query $f$ on dataset $D$, where $Z$ is a random variable drawn from a distribution. If $Z\sim Lap(GS_{f}/\epsilon)$, $A(D)$ provides $\epsilon$-differential privacy. In $\epsilon$-differential privacy, the magnitude of the added noise should be as small as possible to preserve data utility and should be independent of the database for strong privacy protection. Noise calibrated according to global sensitivity is independent from the database $D$ but the magnitude may be too big to make the query results unusable. Noise calibrated according to local sensitivity is dependent on $D$ and makes it fail differential privacy. To address this challenge, Nissim~\textit{et~al.}~\cite{NRS} proposed to use noise calibrated according to the smooth upper bound (more preferably, the smooth sensitivity) of local sensitivity. The basic idea is to add noise proportional to $\frac{S_f(D)}{\alpha}$, that is, $A(D)=f(D)+\frac{S_f(D)}{\alpha}\cdot Z$, where $S_{f}$ is a $\beta$-smooth upper bound on the local sensitivity of $f$,  $Z$ is a random variable with probability density function $h$. 
%
Nissim~\textit{et~al.}~\cite{NRS} pointed out that $h$ must be $(\alpha,\beta)$-\emph{admissible} in order to achieve differential privacy based on smooth sensitivity. 

\begin{definition}[$(\alpha,\beta)$-admissible noise distribution~\cite{NRS}] \label{def:alpha:beta:admissible}
For all $\Delta \in \mathcal{R}$ and $\lambda \in \mathcal{R}$ such that $|\Delta|\leq \alpha$ and $|\lambda|\leq \beta$, a probability density function $h$ is $(\alpha,\beta)$-admissible if it satisfies the following conditions:
\begin{eqnarray}
\mbox{Sliding Property: } & h(z)  \leq  e^{\frac{\epsilon}{2}}\cdot h(z+\Delta)+\frac{\delta}{2} \nonumber \\
\mbox{Dilation Property: } & h(z)  \leq  e^{\frac{\epsilon}{2}}\cdot (e^{\lambda}h(e^{\lambda}\cdot z))+\frac{\delta}{2} \nonumber
\end{eqnarray} 
\end{definition}



The sliding and dilation properties ensure that the noise distribution cannot change much under sliding and dilation, and the values of $\alpha$ and $\beta$ are the upper bounds of $\Delta$ (the sliding offset) and $\lambda$ (the dilation offset) based on $h$. If $h$ of $Z$ is $(\alpha,\beta)$-admissible,the database access mechanism $A(D)=f(D)+\frac{S_f(D)}{\alpha}\cdot Z$ is $(\epsilon, \delta)$-differentially private~\cite{NRS}.

There are three families of admissible distributions: Cauchy, Laplace, and Gaussian~\cite{NRS}. A Cauchy admissible distribution yields a ``pure'' $\epsilon$-differential privacy with $\delta=0$. Laplace and Gaussian admissible distributions can produce an approximate differential privacy with $\delta>0$ with different $\alpha$ and $\beta$ values.


  \begin{figure}[t]
 \centering
  \includegraphics[width=0.8\textwidth]{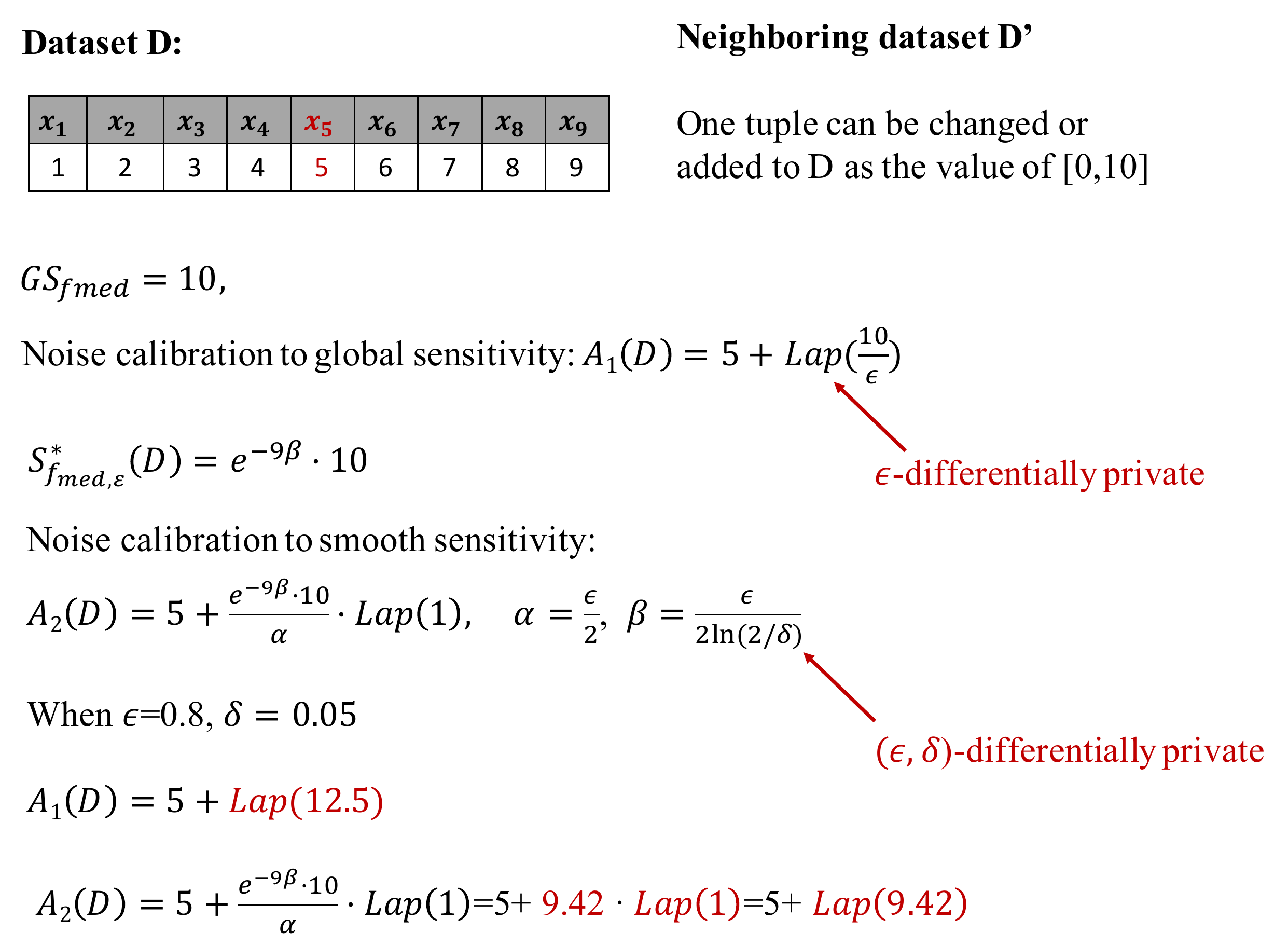}
  \caption{Comparison of noise calibrations.}
  \label{fig:com-sm}
\end{figure}

\begin{figure}[t]
 \centering
  \includegraphics[width=0.8\textwidth]{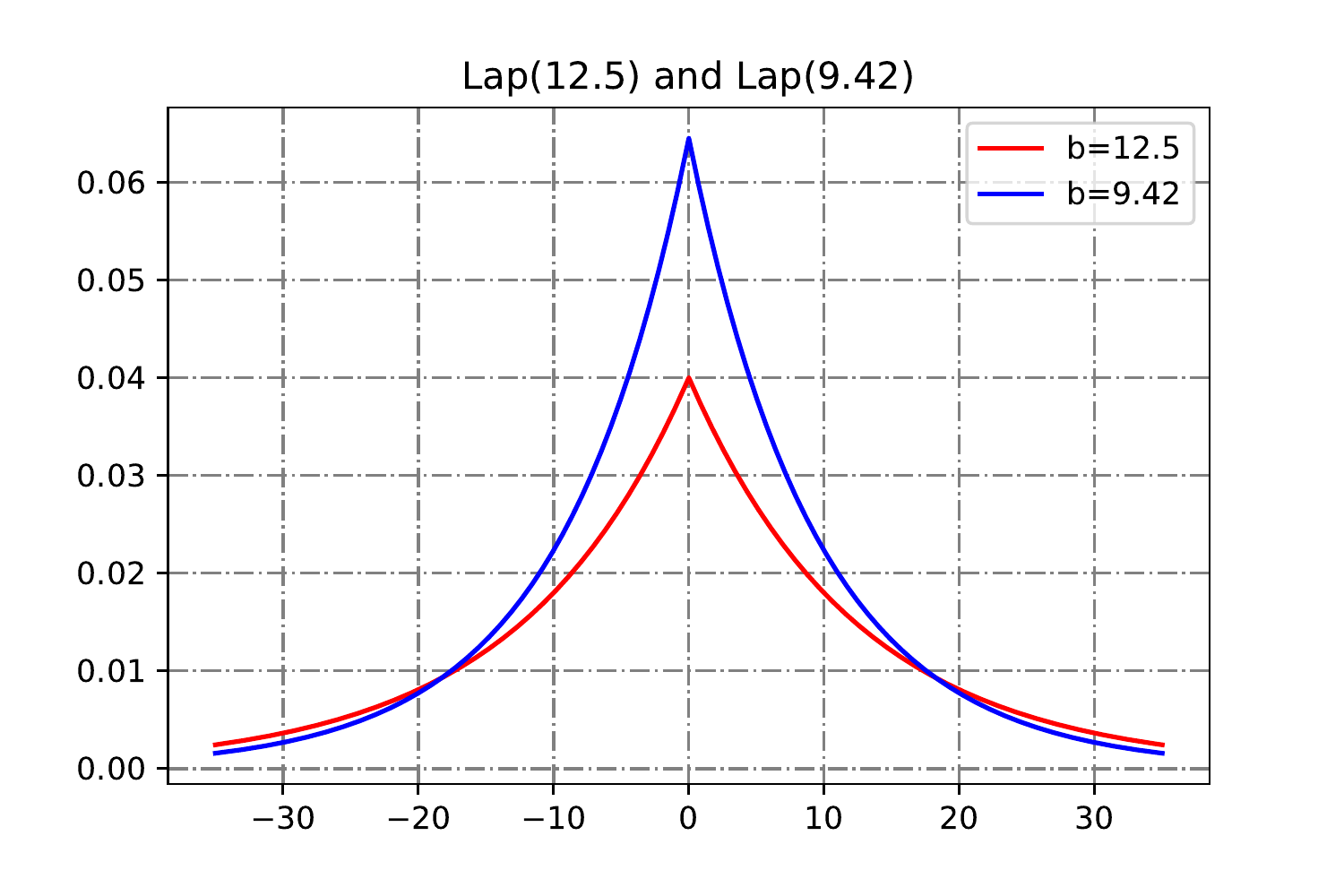}
  \caption{Probability density functions of the two Laplace distributions.}
  \label{fig:lap}
\end{figure}

Consider the \textsf{median} function again and let $D=(x_{1},x_{2},\ldots, x_{n})$, where $x_1\le x_2\le \cdots \le x_n$ and $x_i\in[0,M]$. The global sensitivity of the median function $f$ is $M$. Figure~\ref{fig:com-sm} illustrates the two differentially private mechanisms based on global sensitivity and smooth sensitivity, both utilizing Laplace distributions. The two corresponding probability density functions are shown in Figure~\ref{fig:lap}. The noise calibrated to the smooth sensitivity is less, since the probability of the random variable $Z$ taking a value closer to $0$ is larger, and that of $Z$ being a larger value is smaller. Thus, more noise is added to the output of the global sensitivity based mechanism. In conclusion, for \textsf{median}, at the same privacy protection level (same $\epsilon$), noise calibrated to smooth sensitivity has a smaller magnitude, thus better preserving data utility.
 
\subsection{Composite Differential Privacy}\label{sec:com}
Sometimes a complex privacy preservation problem needs a composite algorithm that involves more than one differential privacy algorithms. More specifically, one may need to sequentially apply various differential privacy algorithms to a dataset, or may need to run various differential privacy algorithms over disjoint datasets to solve a composite problem. The privacy budgets of the composite algorithms for those two cases are summarized by the following two theorems, and the basic concepts are further demonstrated in Figure~\ref{fig:20}.

 \begin{figure}[t]
 \centering
  \includegraphics[width=0.8\textwidth]{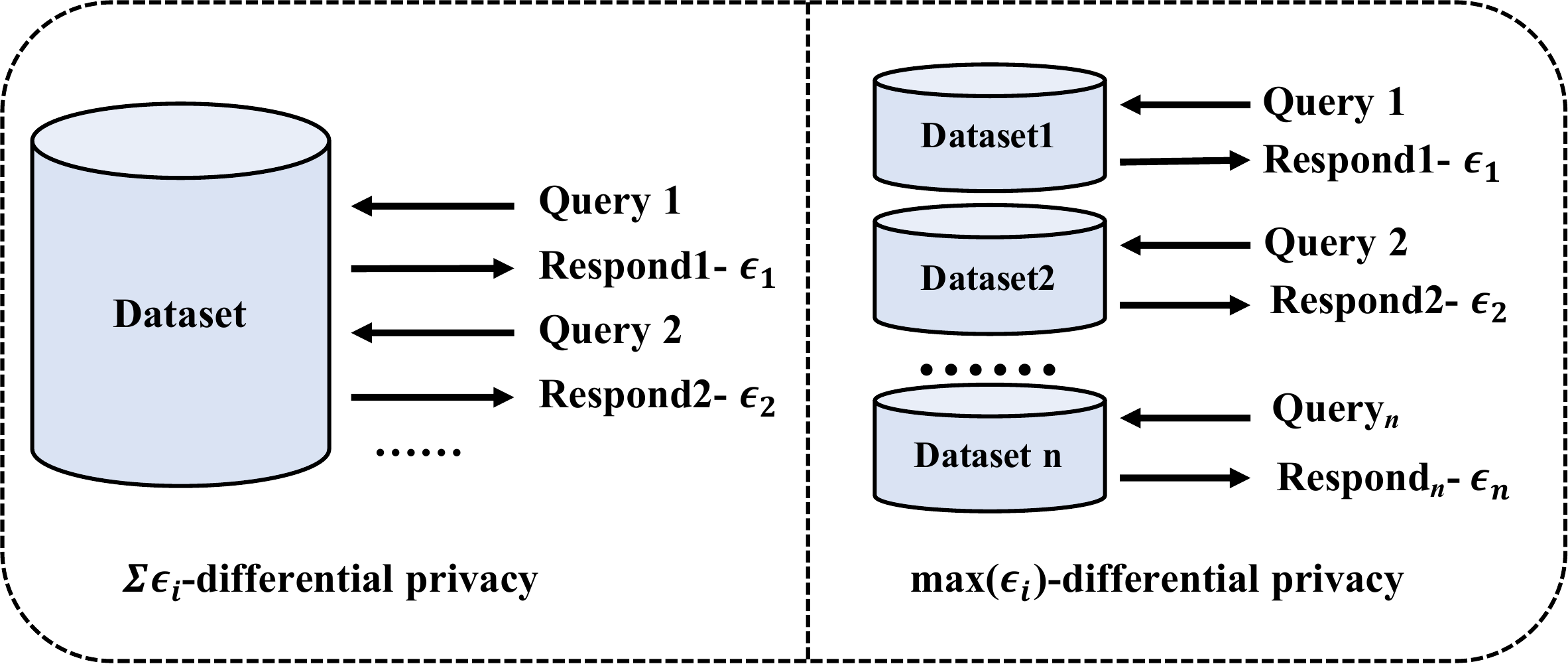}
  \caption{Composition properties of differential privacy.}
  \label{fig:20}
\end{figure}

Let $A_{1}, A_{2},\ldots, A_{n}$ be $n$ $\epsilon$-differential privacy algorithms, whose privacy budgets are respectively denoted by $\epsilon_{1}, \epsilon_{2}, \ldots, \epsilon_{n}$. 

\begin{theorem} \label{theorem:sequential:composition}
(Sequential Composition~\cite{MFP}) The composite algorithm obtained by sequentially applying $A_{1}, A_{2},\ldots, A_{n}$ on the same dataset $D$ provides $\sum\limits_{i=1}^{n}\epsilon_{i}$-differential privacy.
\end{theorem}

\begin{theorem} \label{theorem:parallel:composition}
(Parallel Composition~\cite{MFP}) Let $D_{1}, D_{2},\ldots, D_{n}$ be $n$ arbitrary disjoint datasets. The composite algorithm obtained by applying each $A_i$ on a corresponding $D_i$ provides $\max\{\epsilon_{i}\}$-differential privacy.
\end{theorem}

The above two theorems provide the so-called ``sequential composition'' and ``parallel composition'' properties. Theorem~\ref{theorem:sequential:composition} states the sequential composition property: the level of privacy preservation provided by a composite algorithm consisting of a sequence of differential privacy algorithms over the same dataset is determined by the sum of the individual privacy budgets. Theorem~\ref{theorem:parallel:composition} presents the ``parallel composition'' property: when differential privacy algorithms are applied to disjoint datasets, the overall level of privacy preservation provided by the composite algorithm depends on the worst privacy guarantee among all the differential privacy algorithms, that is, the one with the largest privacy budget. These two theorems can be used to determine whether a composite algorithm satisfies the differential privacy requirement and to reasonably control the allocation of the total privacy budget to each algorithm.

\section{Two Variations of Differential Privacy} \label{sec:var}

To adapt to various problem domains and settings, different variations of differential privacy have been developed. 
In this section, we introduce two most popular variants.  Dependent differential privacy is to handle queries involving correlated database tuples. Local differential privacy is to handle the scenarios where an untrustworthy third-party is employed to collect data.

\subsection{Dependent Differential Privacy}

Differential privacy assumes that the tuples in a database are independent from each other. This assumption is not always true in practice. As indicated by Kifer and Machanavajjhala~\cite{DKA}, the correlation or dependence between tuples may undermine the privacy guarantees of differential privacy mechanisms. Consider a simple example~\cite{LCM}. Let $D=(x_{1},x_{2})$ be a database, and tuples $x_{1}$ and $x_{2}$ have a probabilistic dependence of $x_{2}=0.5x_{1}+0.5Y$, $x_{1}$ and $Y$ have uniform and independent distributions over $[0,1]$, and $Y$ is a random variable to model the relationship between $x_1$ and $x_2$ and to keep $x_1$ and $x_2$ over $[0,1]$; thus we can obtain the global sensitivity 1. Assume that the Laplace mechanism is applied to the \textsf{sum} query $f(D)=x_{1}+x_{2}$. One can see from Figure~\ref{fig:dependent} that the privacy guarantee is $\exp(1.5\epsilon)$ when $x_{1}$ and $x_{2}$ are correlated while it is $\exp(\epsilon)$ if we assume that $x_{1}$ and $x_{2}$ are independent.

\begin{figure}[t]
 \centering
  \includegraphics[width=0.8\textwidth]{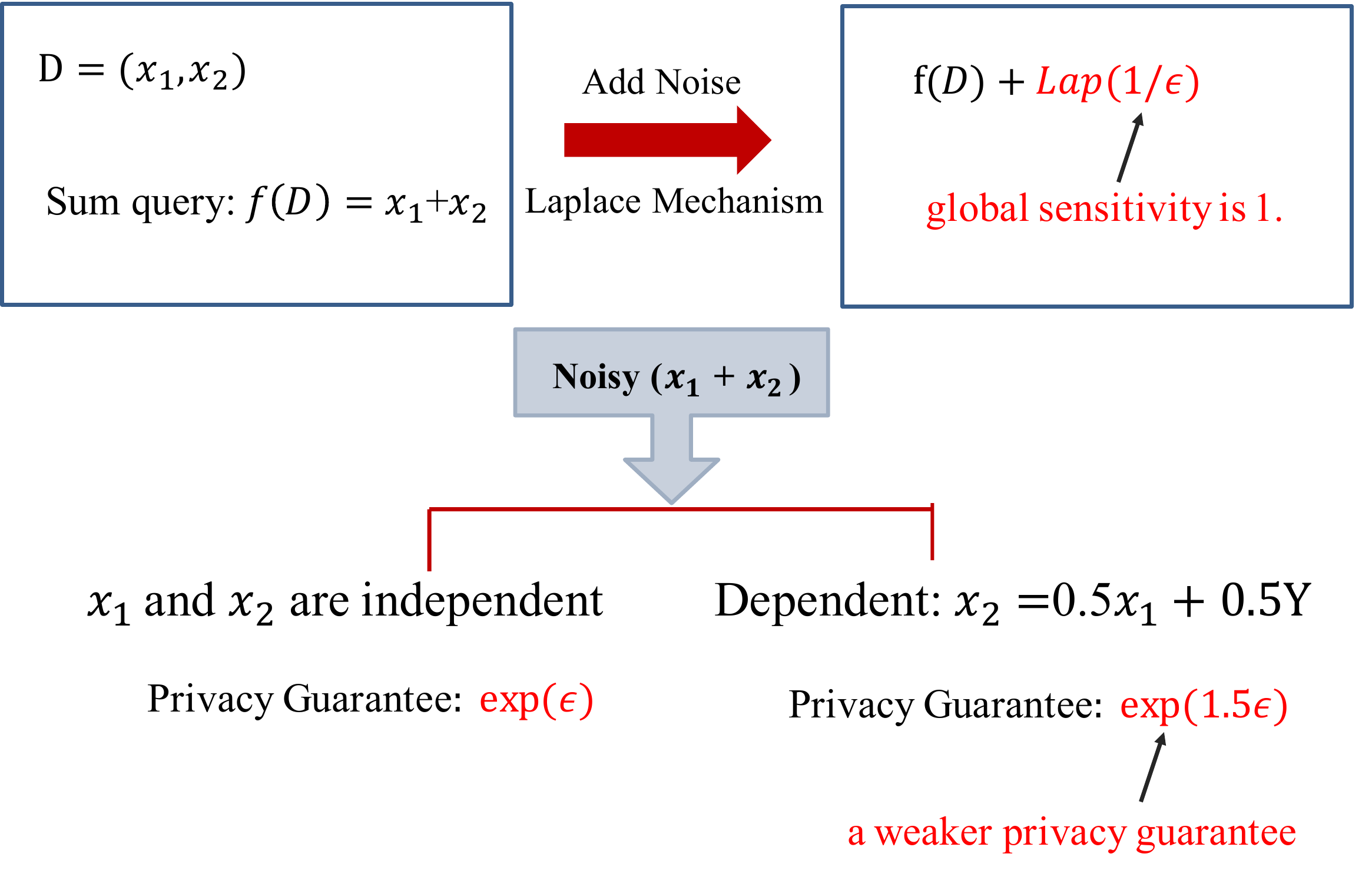}
  \caption{Privacy guarantee for a dependent dataset.}
  \label{fig:dependent}
\end{figure}

\subsubsection{Definition of Dependent Differential Privacy}

In a database $D$, if any tuple is dependent on at most $L-1$ other tuples, the dependence size is $L$. Denote by $R$ the \emph{probabilistic dependence relationship} over the $L$ dependent tuples. 
%
%
Two datasets $D(L,R)$, $D'(L,R)$ are dependently neighboring if changing one tuple in $D(L,R)$ can impact at most $L-1$ other tuples in $D'(L,R)$, where $R$ is the probabilistic dependence relationship among the dependent tuples.

 
\begin{definition}[Dependent Differential Privacy~\cite{LCM}] \label{dependent dp}
 A randomized algorithm $A$ is $\epsilon$-dependent differentially private if for any two dependent neighboring datasets $D(L,R)$ and $D'(L,R)$, and for all sets $S$ of possible outputs, we have
$$\max\limits_{D(L,R),D'(L,R)} \frac{P(A(D(L,R))=S)}{P(A(D'(L,R))=S)}\leq \exp(\epsilon)$$
%
\end{definition}

Dependent differential privacy limits the capacity of an adversary to infer sensitive information, and thus can defend against all possible adversarial inferences even if the adversary has full knowledge of the tuple correlations.

\subsubsection{Laplace Mechanism}

A Laplace mechanism achieving $\epsilon$-dependent differential privacy for a dataset $D(L,R)$ with dependence size $L$ was proposed~\cite{LCM}. For a dataset $D$ with dependence size $L$ and a query function $f$ with global sensitivity $GS_f$, the Laplace mechanism $A(D)=f(D)+Lap(L\cdot GS_f/\epsilon)$ is $\epsilon/L$-differentially private.

Consider the example shown in Figure~\ref{fig:dependent}. The global sensitivity for the \textsf{sum} query is $1$ and the dependence size $L=2$. Thus, the output for this Laplace mechanism is $A(D)=f(D)+Lap(2/\epsilon)$. However, this mechanism implies that all the dependent tuples are \emph{completely dependent} on each other, which makes the query sensitivity $L\cdot GS_f=2$, while the sensitivity of the \textsf{sum} query for the two dependent tuples is $1.5$. In real world datasets, there may be very few tuples that are completely dependent on each other, though they may be related. Thus, many mechanisms consider a fine-grained dependence relationship between tuples to obtain a small dependent sensitivity of queries. For example, Zhao~\emph{et~al.}~\cite{ZZP} adopted the probability graphical models to represent the dependency structure of tuples and achieved high utility.

\subsection{Local Differential Privacy}\label{sec:ldp}

The basic differential privacy setup relies on a trusted third party to collect data, add carefully crafted noise to a query result according to the specification of differential privacy, and publish the noisy statistical results. 
Nevertheless, in practice it is often difficult to find a truly trusted third party to collect and process data.  The lack of trusted third parties greatly limits the applications of the basic, centralized differential privacy. To address this issue, \emph{local differential privacy}~\cite{DJW} emerges, which does not assume the existence of any trusted third-party data collector. 
Instead, it transfers the process of data privacy protection to individual users by asking each of them to independently deal with and protect personal sensitive information. Figure~\ref{fig:21} shows the framework of local differential privacy.
One can see that local differential privacy 
extends its centralized counterpart by 
localizing perturbed data to resist privacy attacks from untrusted third-party data collectors.

\begin{figure}[t]
 \centering{}
  \includegraphics[width=0.8\textwidth]{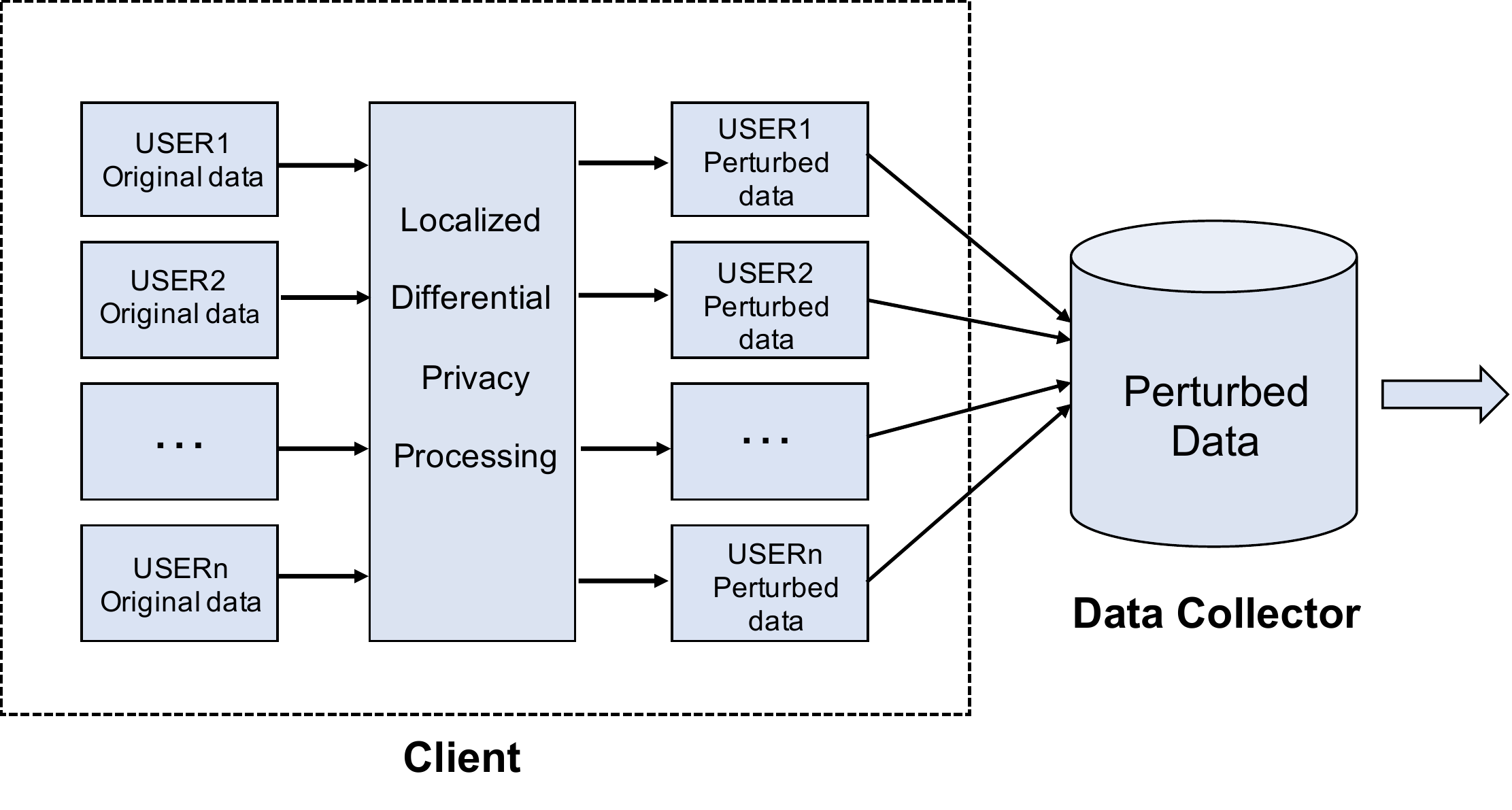}
  \caption{A framework of local differential privacy.}
  \label{fig:21}
\end{figure}

\subsubsection{The Definition of Local Differential Privacy}

\begin{definition}[Local Differential Privacy~\cite{DJW}]\label{ldp}
Consider $n$ users, with each possessing one record. A randomized algorithm $A$ with input and output domains $Dom(A)$ and $Ran(A)$, respectively, is said to satisfy $\epsilon$-local differential privacy if the probability of $A$ obtaining the same output result $t^{*}$ $(t^{*}\subseteq Ran(A))$ on any two records $t$ and $t'$ $(t, t'\in Dom(A))$ satisfies
$$Pr[A(t)=t^{*}]\leq e^{\epsilon}\times Pr[A(t')=t^{*}]$$
\end{definition}

Local differential privacy ensures the similarity between the output results of any two records. By this way it is almost impossible to infer which record is the input data according to an output result of algorithm $A$. In centralized differential privacy, the privacy guarantee of algorithm $A$ is defined on neighboring datasets, and requires a trusted third-party data collector. Nevertheless, in local differential privacy, each user processes its individual data independently, that is, the privacy preserving process is transferred from the data collector to individual users, so that a trusted third party is no longer needed and privacy attacks brought from an untrusted third-party data collector is thus avoided. 
The implementation of local differential privacy requires data perturbation mechanisms.

\subsubsection{Perturbation Mechanisms}

The random response technique~\cite{WSL} proposed by Warner in 1965 is the mainstream perturbation mechanism adopted by local differential privacy. 
The main idea is to protect data privacy by making use of the uncertainty in the responses to sensitive questions. Consider an example of $n$ persons with an unknown  proportion $\pi$ of diseased patients. To calculate $\pi$, a survey question is launched: "are you a patient with some disease?" Each user responds with either ``Yes'' or ``No''. For privacy preservation, a user may not respond with the true answer. Assume that a user responds with the help of a non-uniform coin flip in which the probability of heads showing up is $p$ and the probability of tails showing up is $1-p$. Then if a head shows up, the user responds with the true answer; otherwise, it responds with the opposite.

\nop{, as shown in Figure~\ref{fig:warner}. 

\begin{figure}[t]{}
 \centering
  \includegraphics[width=0.4\textwidth]{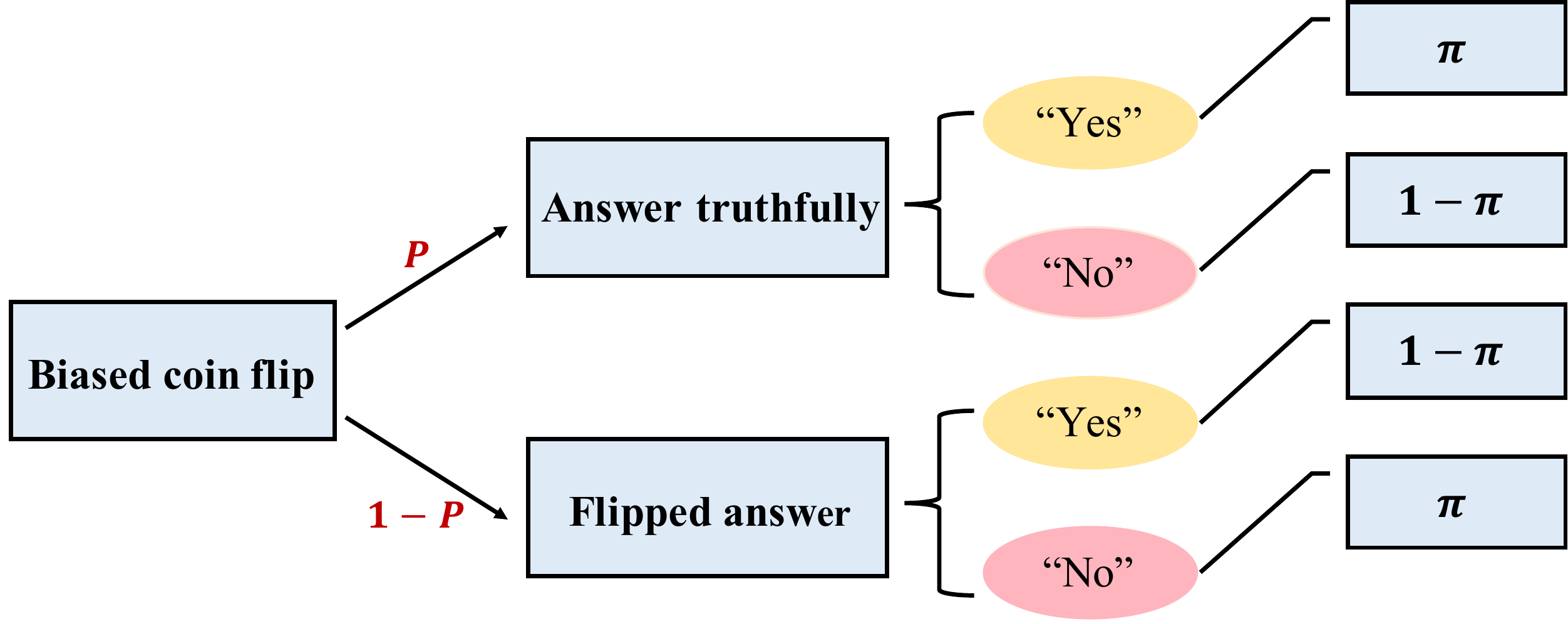}
  \caption{Diagram of the Warner model.}
  \label{fig:warner}
\end{figure}
}

The mechanism achieves $\epsilon$-local differential privacy, where
$\epsilon=|\ln\frac{p}{1-p}|$.  Consider the concrete example in Figure~\ref{fig:ldp}, where each individual randomly responds the survey question with a biased coin flip of $p=3/4$. The data collector aggregates all responses from the users and estimates the count of diseased persons. This mechanism achieves $\ln3$ local differential privacy.
\begin{figure}[t]{}
 \centering
  \includegraphics[width=0.8\textwidth]{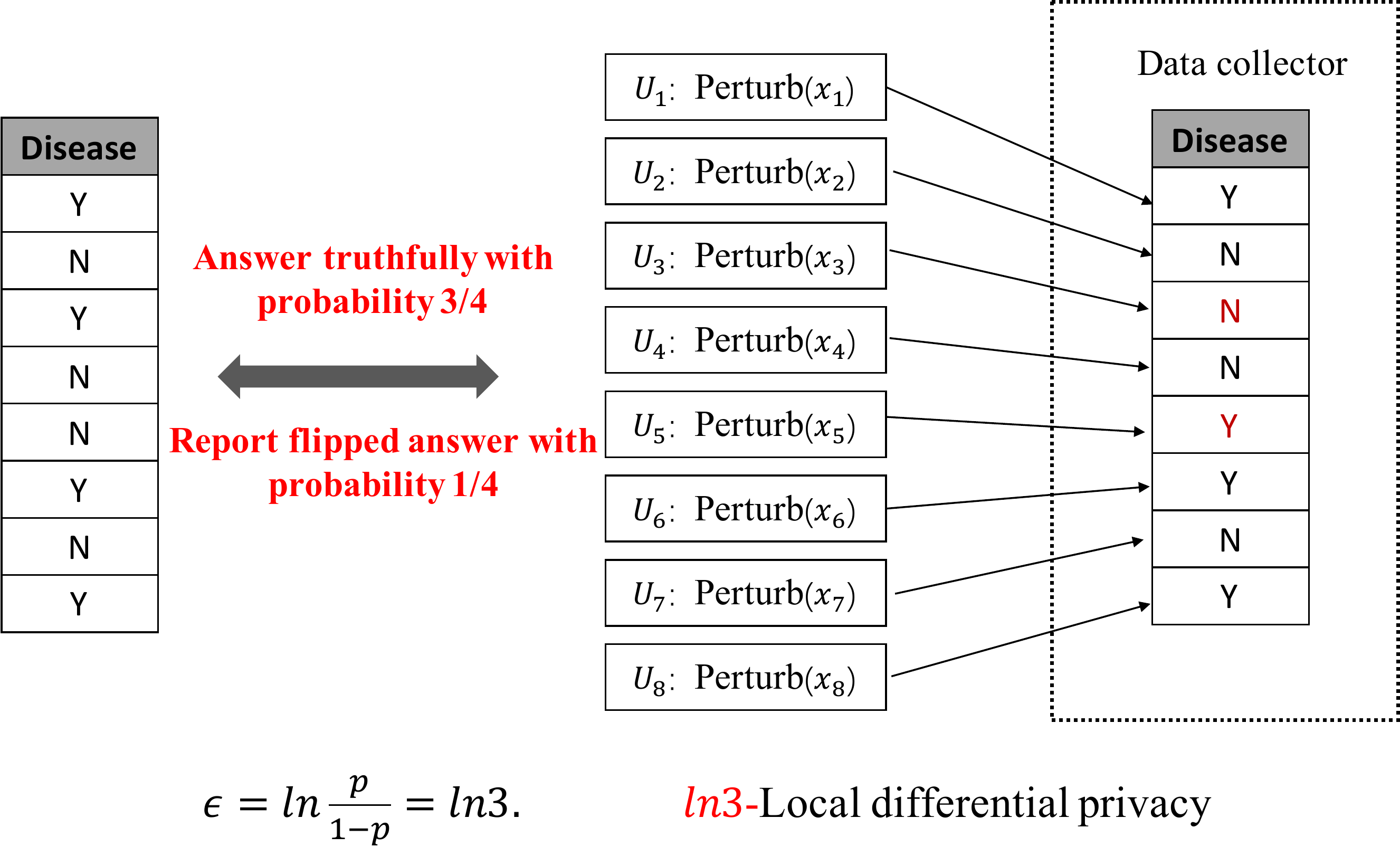}
  \caption{An example of LDP mechanism.}
  \label{fig:ldp}
\end{figure}

The Warner model mentioned above is simple and influential. Some variations and extensions are developed, such as the \emph{Mangat Model}~\cite{mangat1994} and the \emph{forced alternative response}~\cite{force}. 
Some other perturbation mechanisms such as \emph{information compression} and \emph{distortion} were also employed by different applications~\cite{XSM,SSL}.

\subsubsection{Composition}
As mentioned in Section~\ref{sec:com}, sequential composition and parallel composition are employed to provide a differentially private solution to a complex problem that involves more than one queries. Sequential composition emphasizes that the privacy budget can be allocated in different steps of an algorithm, while parallel combination guarantees the privacy of the algorithm satisfying differential privacy on the disjoint subsets of a database. By definition, centralized differential privacy is based on ``neighboring datasets'' and local differential privacy is defined on any two records of a dataset. The forms of privacy guarantee are the same. Therefore, local differential privacy inherits the sequential and parallel composition features mentioned in Section~\ref{sec:com}.

\section{Privacy Attacks and Types of Differential Privacy for Social Networks}\label{sec:dpg}

In this section, we first summarize the popular privacy attacks and provide insights on how to model privacy in social networks. Then, to adapt differential privacy from tabular data to social network data, we present the types of differential privacy, including node privacy, edge privacy, out-link privacy and partition privacy.
 
\subsection{Privacy Attacks in Social Networks}

Privacy attacks~\cite{zheleva2011,ZBPL,AVS,AVS2,DRT2012,GNT2014} refer to a wide variety of activities that leak sensitive information to unauthorized parties who should not know the information. The most serious type of privacy attacks in online social networks is inference attacks~\cite{ASHF}, which breach users' private information by analyzing background knowledge, such as user occupations or salary. Two classes of inference attacks are observed in social networks, namely private attribute inference~\cite{DRT2012,GNT2014,LSW2013,MAV2010,WUBS,CAAG,KMSD} and user de-anonymization~\cite{AVS,AVS2,GTE,SWN,SWM2,JXC,JTJ,SGE,SLS}. 

Private attribute inference aims to reveal a hidden attribute value that is intentionally protected by the user or service provider. Neighbor-based inference attacks~\cite{DRT2012,GNT2014,LSW2013,MAV2010} abuse the fact that adjacent users may have the same or similar attribute values with a high probability and infer the private attribute of one user by exploiting the known attribute values of some other users sharing similar interests~\cite{CAK2012}. For example, if the majors of more than half of a user's friends are ``computer science'', then the user has a high probability of majoring in ``computer science''.  Behavior-based inference~\cite{WUBS,CAAG,KMSD} tries to identify the similarities of certain attribute values through the behavioral data, such as interests, characteristics and cultural behaviors. For example, if most of the apps, books, and musics that a user likes are from China, there is a high probability that the user was originally from China.

User de-anonymization~\cite{AVS,AVS2,GTE,SWN,SWM2,JXC,JTJ,SGE,SLS} takes an anonymized graph and a reference graph having the true user identities as inputs and maps the nodes in these two graphs such that the identities of the users in the anonymized graph can be reidentified. An anonymized social network graph is usually released by a service provider to various requesters, such as researchers, advertisers, application developers and government agencies, after hiding private identifiable information by various anonymization techniques, such as pseudonyms, graph modification, clustering and generalization. A reference graph can be easily obtained through the gathered information from other sources such as a different social network which has overlapping users with a published social graph.  Typically, a reference graph may have less attributes about nodes than an anonymized social network graph.

These two categories of privacy attacks lead to the exposure of different sensitive information. To protect private data in a social network and formalize the notion of ``privacy'' in social networks, distinctive privacy threats are recognized, which are shown in Figure~\ref{fig:attack}.

 \begin{figure}[t]
 \centering
  \includegraphics[width=0.8\textwidth]{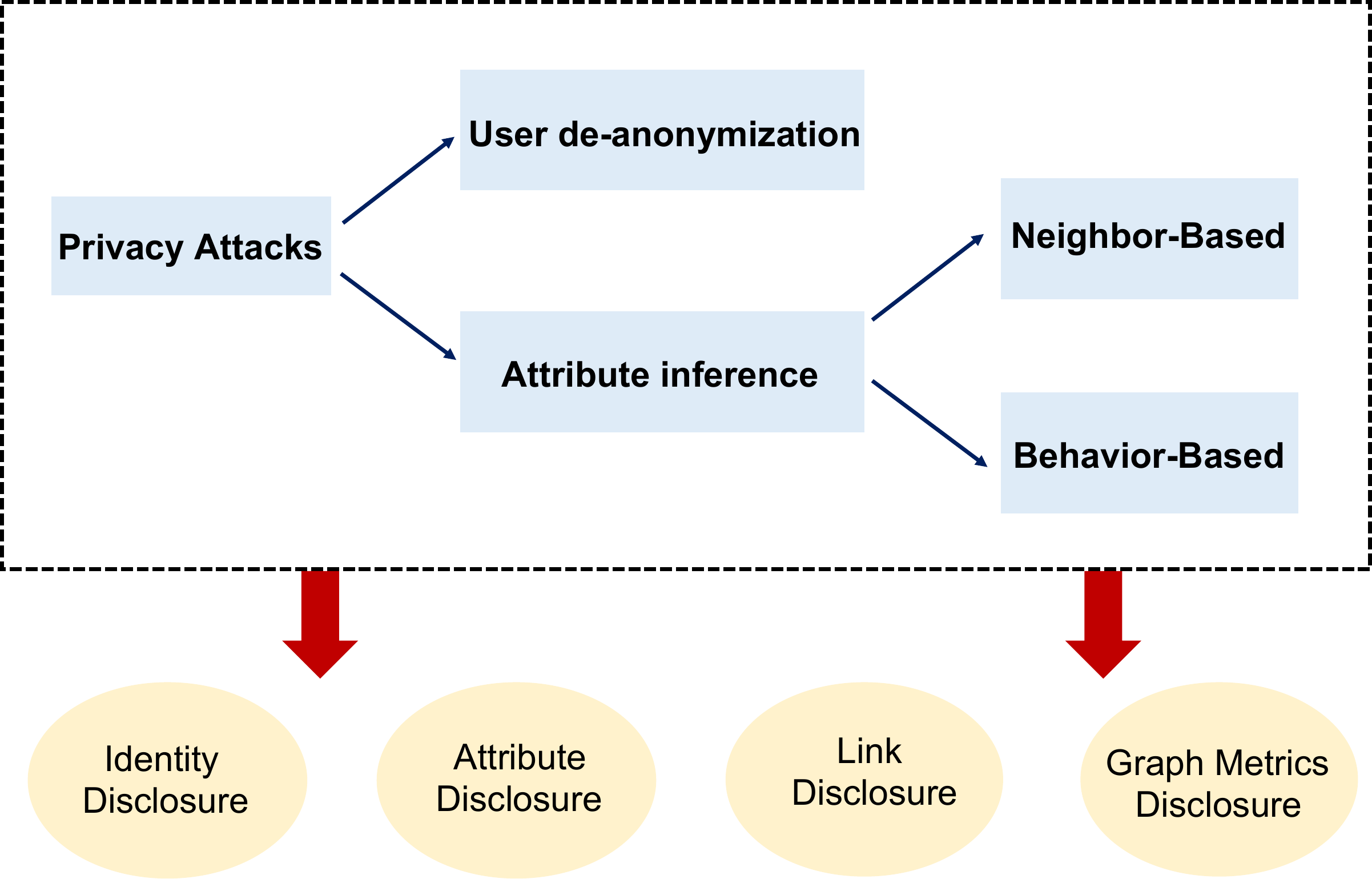}
  \caption{Privacy attacks in online social networks.}
  \label{fig:attack}
\end{figure}

\begin{itemize}
\item \textbf{Identity disclosure~\cite{zheleva2011,KMMN}:} In social networks, the identity of an individual may be considered private, while attackers may exploit various user information to reidentify a social network user or to determine whether or not a target individual is present in a social network. For instance, AOL released an anonymized partial three-month search history to the public in 2006. Although personally identifiable information was carefully processed, some identities were accurately reidentified -- \emph{The New York Times} immediately located the following individual: the person with number $4417749$ was a $62$-year-old widowed woman who suffered from some diseases and had three dogs.

\item \textbf{Attribute disclosure~\cite{zheleva2011,BLDC,HMMG,LKTE,ZBPL}:} A social network user's profile usually includes various attributes such as age, gender, major and occupation, some of which, such as salary, health status and disease information, are considered sensitive and private.

\item \textbf{Link disclosure~\cite{KMMN,ZBPL,zheleva2011}:} The social relationships between individuals can be modeled as edges in a social graph. The link information may be considered sensitive in some cases. For example, Kossinets and Watts~\cite{KGW} analyzed a graph derived from email communications among students and faculty members in a university, of which the email relationships of ``who emailed whom'' was deemed sensitive~\cite{HLM}. 

\item \textbf{Graph Metrics disclosure~\cite{ji2016}:} Since social networks can be modeled as graphs, graph metrics, such as degree, betweenness, closeness centrality, shortest path length, subgraph counting and edge weight, may be employed to conduct social network analysis. The disclosure of such information may indirectly lead to privacy leakage. For example, many de-anonymization attacks are based on the structure information of a social graph.

\end{itemize}

Modeling privacy is critical for realizing privacy preservation in social networks. Differential privacy assumes the maximum background knowledge for adversarials. In the subsequent section, we present how to protect a social network under differential privacy by extending the differential privacy definition from traditional databases to graphs, and demonstrate how to formally define differential privacy in social networks based on the privacy threats mentioned above.

\subsection{Types of Differential Privacy for Social Networks}

A social network can be modeled as a graph $G(V,E)$, where $V$ is a set of nodes and $E$ is a set of relational activities between nodes. Differential privacy originates from traditional databases.
The key to extending differential privacy to social networks is to determine the neighboring input entries, that is, how to define ``adjacent graphs''. In this subsection, we review the applications of differential privacy in social networks by instantiating the adjacent graphs into node privacy~\cite{KRS}, edge privacy~\cite{KRS}, out-link privacy~\cite{TCC} and partition privacy~\cite{TCS}. 

\subsubsection{Node Privacy}

A privatized query $Q$ preserves \emph{node privacy}~\cite{HLM} if it satisfies differential privacy for every pair of graphs $G_{1}=(V_{1}, E_{1})$ and $G_{2}=(V_{2}, E_{2})$ such that $|(V_{1}\cup V_{2})\setminus (V_{1}\cap V_{2})|=1$ and  $\{(E_{1}\cup E_{2})\setminus (E_{1}\cap E_{2})\}=
\{(u,v)|u=x\vee v=x\}$, where $x$ is the only node in $(V_{1}\cup V_{2})\setminus (V_{1}\cap V_{2})$ and $(u,v)$ represents the edge between nodes $u$ and $v$.

\nop{
\begin{figure}[t]
 \centering
  \includegraphics[width=0.44\textwidth]{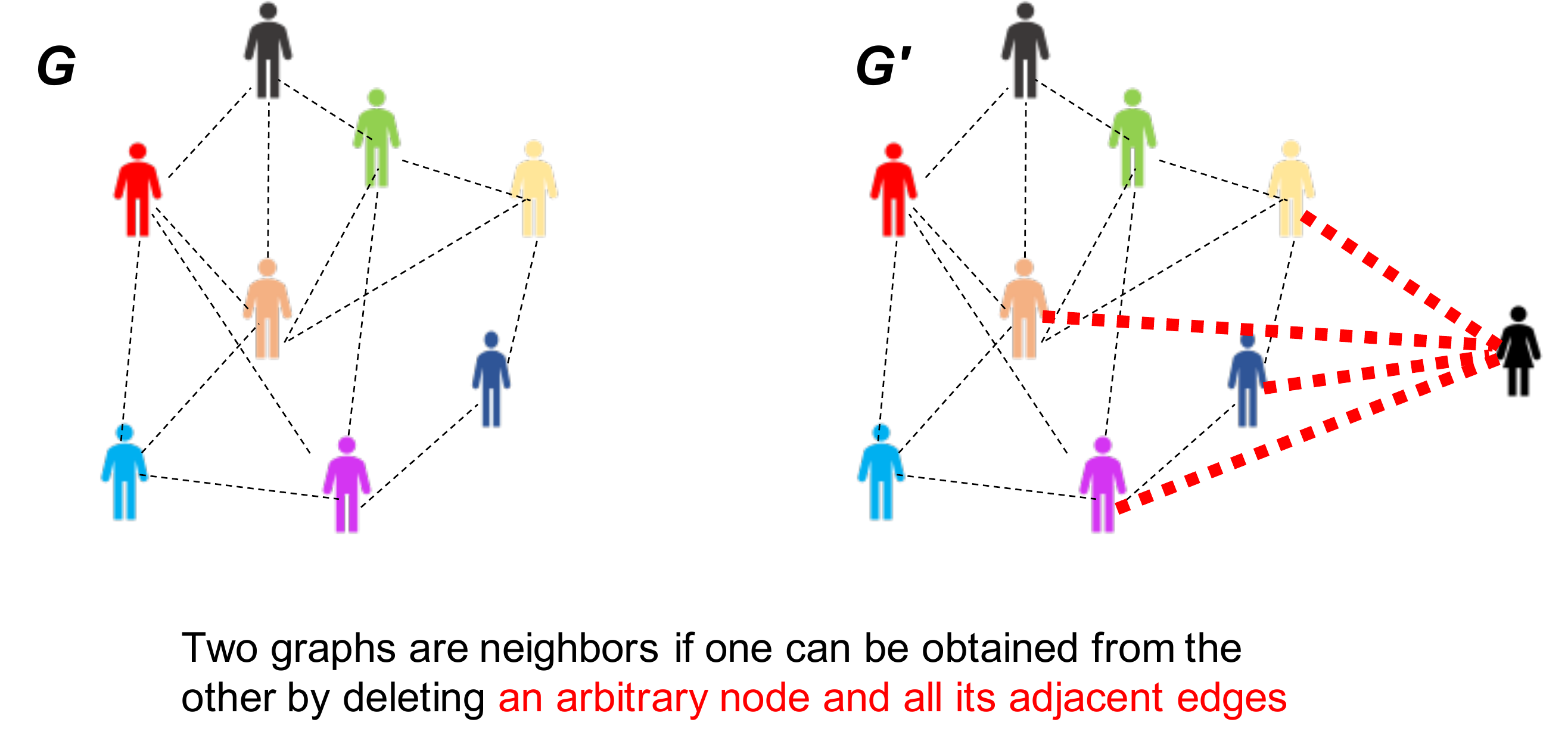}
  \caption{Node privacy.}
  \label{fig:node}
\end{figure}
}

In node privacy, an adjacent graph $G'$ of a given social network $G$ is the one obtained by deleting or adding a node and all edges incident to the node. 
Node differential privacy tries to prevent an attacker from determining whether or not an individual node $x$ appears in the graph. It guarantees privacy preservation for individuals and relationships simultaneously rather than just a single relationship, at the cost of strict restrictions on queries and reduced-accuracy results. 
A differentially private algorithm must conceal the worst-case discrepancy between adjacent graphs, which may be substantial under node privacy. For example, if we consider an extreme case where a node connects to all other nodes (a star graph), then the sensitivity is high and the added noise has to be dramatic, too. Generally speaking, node privacy is infeasible to provide high utility (accurate network analysis) due to high sensitivity, but it provides desirable privacy protection~\cite{HLM}.

\subsubsection{Edge Privacy}

A privatized query $Q$ preserves \emph{edge privacy}~\cite{HLM} if it satisfies differential privacy for each pair of graphs $G_{1}=(V_{1}, E_{1})$ and $G_{2}=(V_{2}, E_{2})$ such that $V_{1}=V_{2}$ and $|(E_{2}\cup E_{1})\setminus(E_{2}\cap E_{1})|=1$. 
\nop{
 \begin{figure}[t]
 \centering
  \includegraphics[width=0.44\textwidth]{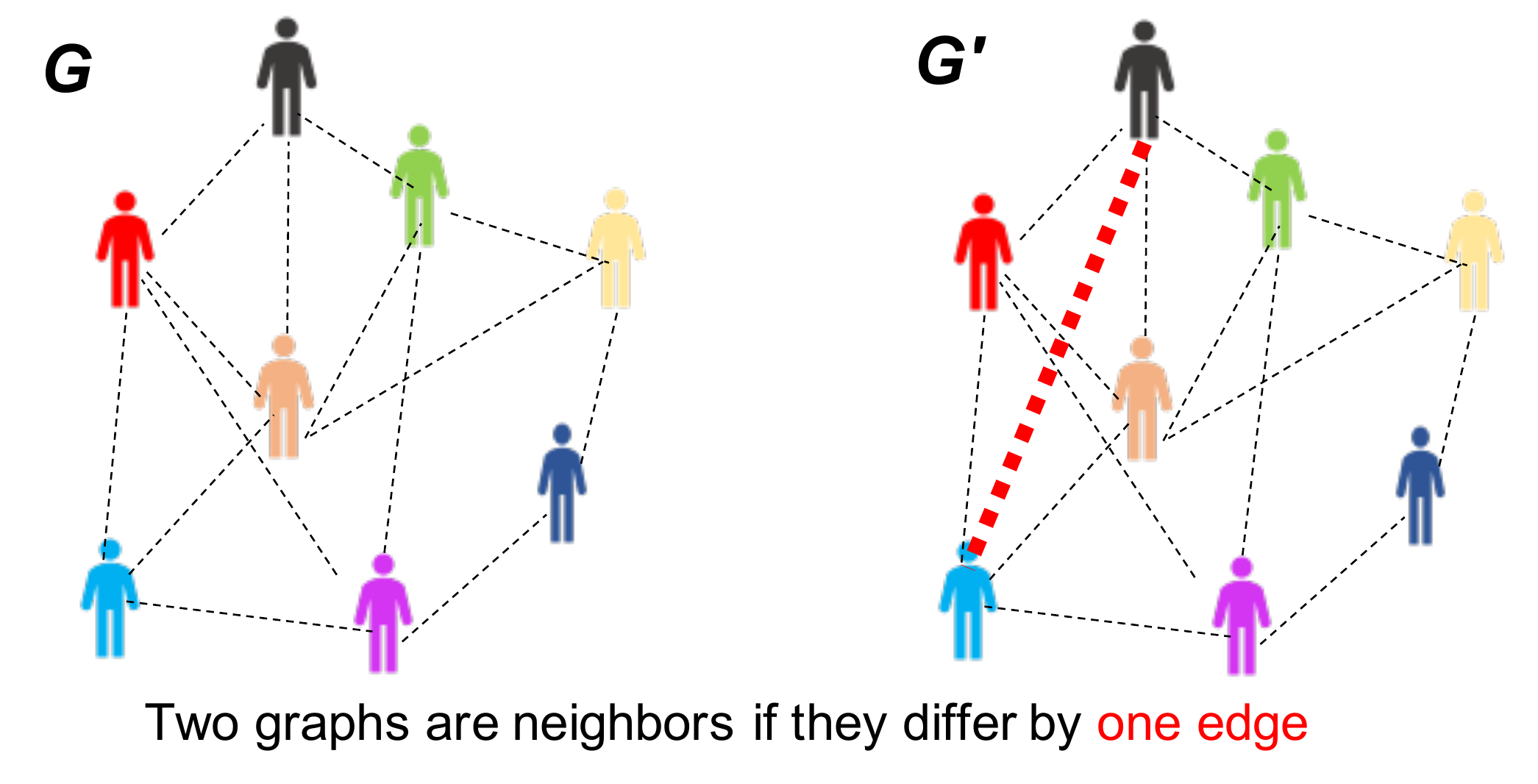}
  \caption{Edge privacy.}
  \label{fig:edge}
\end{figure}
}
In edge privacy, an adjacent graph $G'$ of a given social network $G$ is obtained by deleting or adding one edge from $G$. It can be generalized to allow at most $k$ edges are changed.

Edge privacy protects against learning about specific relationships between users and prevents an attacker from determining with a high certainty whether two individuals are connected. Comparing with node privacy, edge privacy can only provide protection on information about relationships between users thus. Nodes with higher degrees still have a higher impact on query results, despite the fact that the relationships between these nodes have been protected. Edge privacy provides meaningful privacy protection in many practical applications and is more widely used than node privacy~\cite{TCC}. For example, Kossinets and Watts~\cite{KGW} employed edge privacy to protect email relationships.

\subsubsection{Out-Link Privacy}

A privatized query $Q$ preserves \emph{out-link privacy}~\cite{TCC} when it meets the definition of  differential privacy for every pair of graphs $G_{1}=(V_{1}, E_{1})$ and $G_{2}=(V_{2}, E_{2})$ such that $V_1=V_2$ and there exists a node $x$ such that $\{(E_{1}\cup E_{2})\setminus (E_{1}\cap E_{2})\}=\{(x\rightarrow v)|x\in V_1 \wedge v\in V_2 \mbox{ or } x\in V_2 \wedge v\in V_1\}$, where $(x\rightarrow v)$ is a directed link from $x$ to $v$.
\nop{
\begin{figure}[t]
 \centering
  \includegraphics[width=0.44\textwidth]{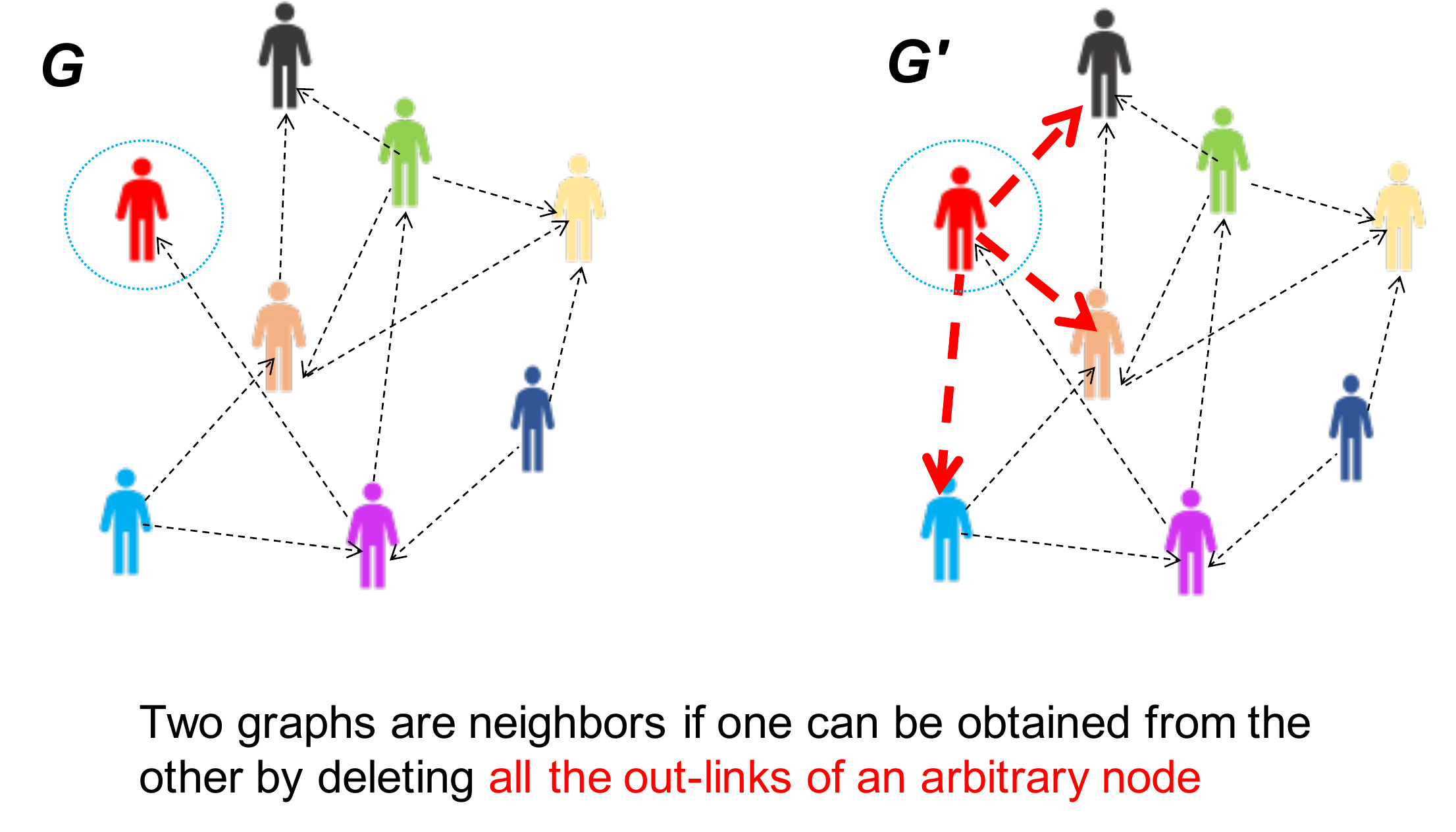}
  \caption{Out-link privacy.}
  \label{fig:out}
\end{figure}
}
In out-link privacy, for a given social network $G$, an adjacent graph $G'$ is obtained by either removing all the existing out-links of a node $x$, or adding one or more new out-links to a node whose out-degree in $G$ is $0$.

Out-link privacy can reduce the distinguishing properties of high-degree nodes, that is, a high-degree node can deny that the friendships are mutual in query results although others claim to be friends with this node. 
Out-link privacy is strictly weaker than node privacy, but for certain query functions it has better performance than edge privacy~\cite{TCS}. Out-link privacy simplifies the calculation of sensitivity and reduces the amount of injected noise required, thus allows certain queries that are infeasible under node privacy and edge privacy~\cite{TCC}. We take the degree distribution as an example and demonstrate that the out-link privacy requires less noise in Section~\ref{sec:degree}.

\subsubsection{Partition Privacy}

A partitioned graph $G$ is comprised of multiple disjoint components $H_i$~\cite{TCS}. A privatized query $Q$ preserves \emph{partition privacy} if it satisfies differential privacy for every pair of graphs $G_{1}$ and $G_{2}$, where $G_{1}=G_{2}-H_{i}$ with $H_{i}\in G_{2}\wedge H_i\notin G_1$ or  $G_{2}=G_{1}-H_{j}$ with $H_{j}\in G_{1}\wedge H_j\notin G_2$.

In partition privacy, an adjacent graph of a given social network $G$ is obtained by adding a new or deleting an existing subgraph from $G$. 
Most social-structure queries are conducted over a set of subgraphs instead of a connected social graph. Some attributes of the nodes such as address, major, and education level can be used to partition a large social graph into multiple subgraphs, and each subgraph can be treated as a multi-attribute data point. Then, deleting or inserting a subgraph is equivalent to removing or adding a data point~\cite{TCS}. Accordingly, traditional differential privacy can be applied to the set of subgraphs (data points). 

Partition privacy provides broader preservation than node privacy, and the protection is applied not to a single node, but to a social group.

\nop{
 \begin{figure}[t]
 \centering
  \includegraphics[width=0.44\textwidth]{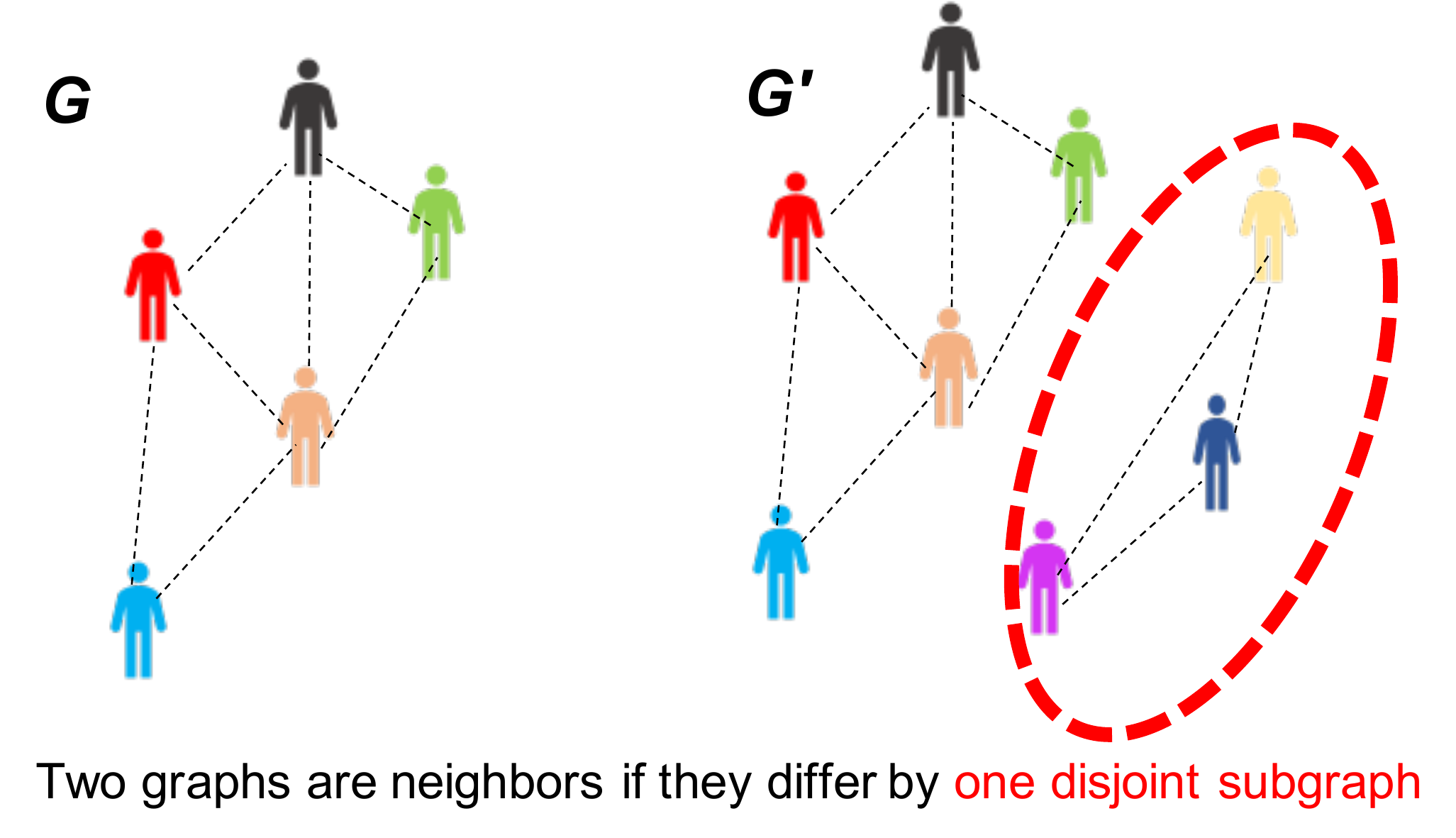}
  \caption{Partition privacy.}
  \label{fig:part}
\end{figure}
}


\section{Differential Privacy in Social Network Analysis}\label{sec:snat}

In this section, we summarize the state-of-the-art research on a series of most-widely used differentially private social network analysis techniques. Social network analysis refers to the quantitative analysis on the data generated by social network services using statistics, graph theory and other techniques. 
Some popular tasks of social network analysis include degree distribution, triangle counting, $k$-star counting, $k$-triangle counting and edge weight analysis. In this section, we analyze a few widely used techniques in social network analysis under differential privacy preservation. Table~\ref{tab:graph} summarizes the major existing differentially private social network analysis techniques for degree distribution and subgraph counting while those for edge weight is summed up in Table~\ref{tab:5}.

\begin{sidewaystable}[]
\renewcommand{\arraystretch}{1.2}

\caption{Summary on Existing Differentially Private Graph Analysis Techniques} 
\centering
\scalebox{0.9}{
\begin{tabular}{ccccc}
\hline
\multicolumn{1}{c}{\multirow{2}{*}{\textbf{Graph statistics}}} & \multicolumn{4}{c}{\textbf{Privacy standard}}                                                                                                                                                                                                                                                                                                                                                                                     
\\ \cline{2-5} 
\multicolumn{1}{c}{}   & \textbf{Edge privacy} & \textbf{Node privacy} & \textbf{Out-link Privacy} & \textbf{Partition Privacy}\\
\hline
\hline
Degree distribution    & ~\cite{HLM}, ~\cite{hrms} & 
\begin{tabular}[c]{@{}l@{}}\cite{knrs},~\cite{dll},~\cite{MPSJ}, \\\cite{slmv},~\cite{rs},~\cite{SUJ} \end{tabular}  &~\cite{TCS}  &   ~\cite{TCS}                                                                                                          \\ 

\hline

Graph publishing/sharing   & ~\cite{ALJ},~\cite{SZW} &       &              &                                                                                                                   \\ 
\hline
Triangle counting, Centrality  &           &        &   ~\cite{TCS}     &  ~\cite{TCS}                                                                                                                 \\ 
\hline
\begin{tabular}[c]{@{}l@{}}Triangle counting, \\Clustering, MST cost \end{tabular}  &~\cite{NRS}  &        &         &                                                                                                                        \\ 

\hline
Cut function of graph   &~\cite{gru}          &        &         &                                                                                                                 \\ 

\hline
Subgraph counting  & \begin{tabular}[c]{@{}c@{}}\cite{KRS},~\cite{zcp},~\cite{BBD},\\~\cite{rhms} (Weaker than edge privacy) \end{tabular}            & ~\cite{knrs},~\cite{slmv},~\cite{BBD}        &              &                                                                                                       \\ 
\hline

\begin{tabular}[c]{@{}c@{}}Average degree, \\ Distance to connectivity\end{tabular} &          & \begin{tabular}[c]{@{}c@{}}~\cite{glp} \\(Stronger than node privacy)  \end{tabular}       &       &                 
                                                                                   \\ 
\hline

Graph generation   & \begin{tabular}[c]{@{}c@{}}~\cite{ZTY}\\Edge-LDP \end{tabular}  &        &         &                                                                                                                 \\ 
\hline

Triangle counting, $k$-clique  & \begin{tabular}[c]{@{}c@{}}~\cite{SXK} \\(Stronger than Edge-LDP) \end{tabular} &        &         &                                                                                      \\ 
\hline
\end{tabular}}
\label{tab:graph}
\end{sidewaystable}

\subsection{Degree Distribution}\label{sec:degree}

Degree distribution is one of the most widely studied graph characteristics. It reflects the graph structure statistics and may affect the whole process of graph operations. Degree distributions can be employed to describe the basic social network structures, design graph models and measure graph similarities.

The degree distribution of a graph can be simply transformed to a degree sequence by counting the frequency of each degree. Here we use a degree histogram to describe the degrees of the nodes in a graph. Consider the example shown in Figure~\ref{fig:9}. One can see that the degree counts change significantly when deleting node $A$. This implies that the sensitivity of degree distribution is high under node privacy since the change of one node may affect multiple degree counts. A careful analysis reveals that a node of degree $k$ affects $2k+1$ values of the histogram at most. In the worst case, the addition or deletion of a node of the maximum degree results in the change of $2n+1$ values, which indicates that the global sensitivity depends on the value of $n$, the number of nodes in the graph. Since $n$ is unbounded, the degree histogram (distribution) query is not feasible for differential privacy protection under node privacy.

\begin{figure}[t]
 \centering
  \includegraphics[width=0.8\textwidth]{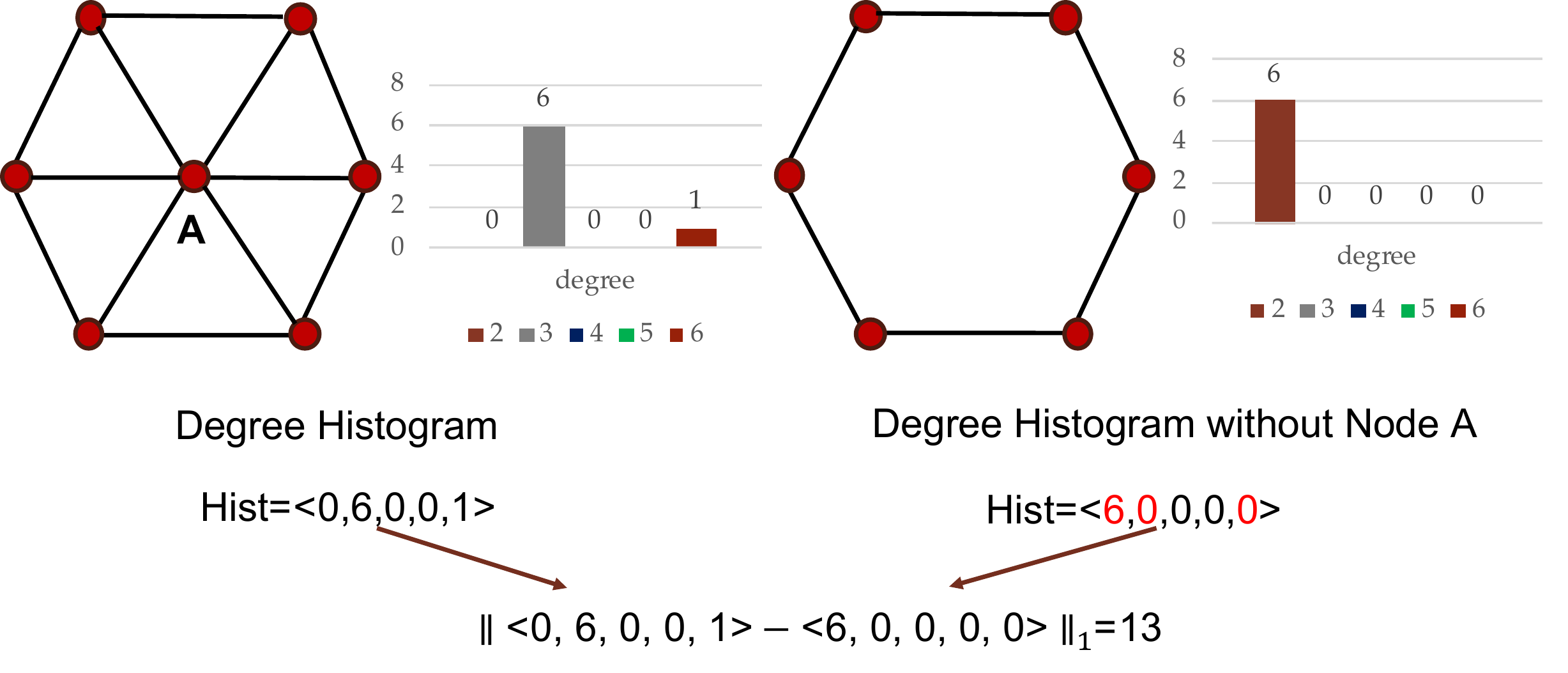}
  \caption{Degree histogram under node privacy.}
  \label{fig:9}
\end{figure}

Under edge privacy, protecting degree histogram queries using differential privacy is feasible, as illustrated in Figure~\ref{fig:10}. One can see that removing an edge from a network only changes the degrees of two nodes, thus affecting $4$ counts at most. The sensitivity is $4k$ under the $k$-edge privacy. Accordingly, when $k$ is small, the amount of added noise is relatively small and even negligible for a graph that is large enough, providing preservation in data utility.

\begin{figure}[t]
 \centering
  \includegraphics[width=0.8\textwidth]{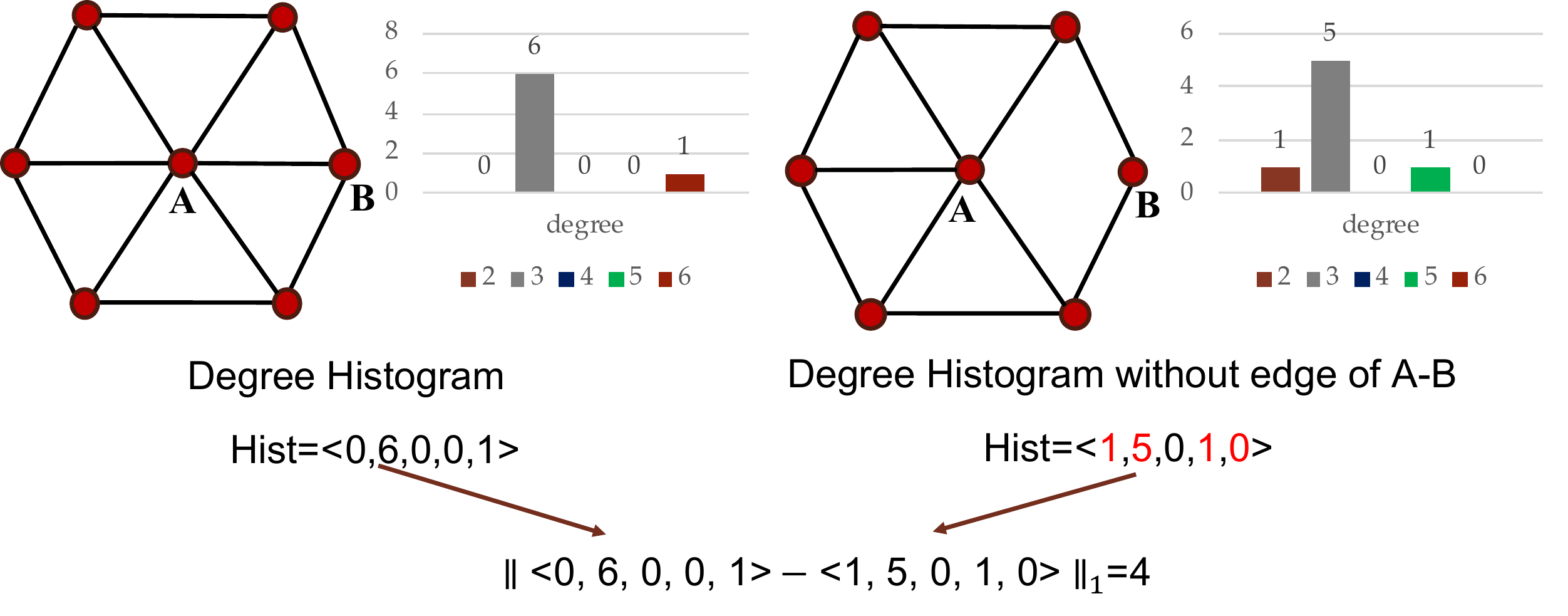}
  \caption{Degree histogram under edge privacy.}
  \label{fig:10}
\end{figure}

Out-link privacy requires less noise for a degree histogram query. Removing out-links of one node from a graph affects one value in the histogram when only out-degrees are counted, as shown in Figure~\ref{fig:11}. Under the out-link privacy, a high-degree node may be identified according to its neighbors' out-degrees. Nevertheless, a slightly higher-than-expected node degree in a graph may not be easily identified~\cite{TCS}. Therefore, if an attacker intends to guess the presence of a high-degree node with certainty, she may have to learn full knowledge about the social network. Thus, out-link privacy improves edge privacy to some extent.

\begin{figure}[t]
 \centering
  \includegraphics[width=0.8\textwidth]{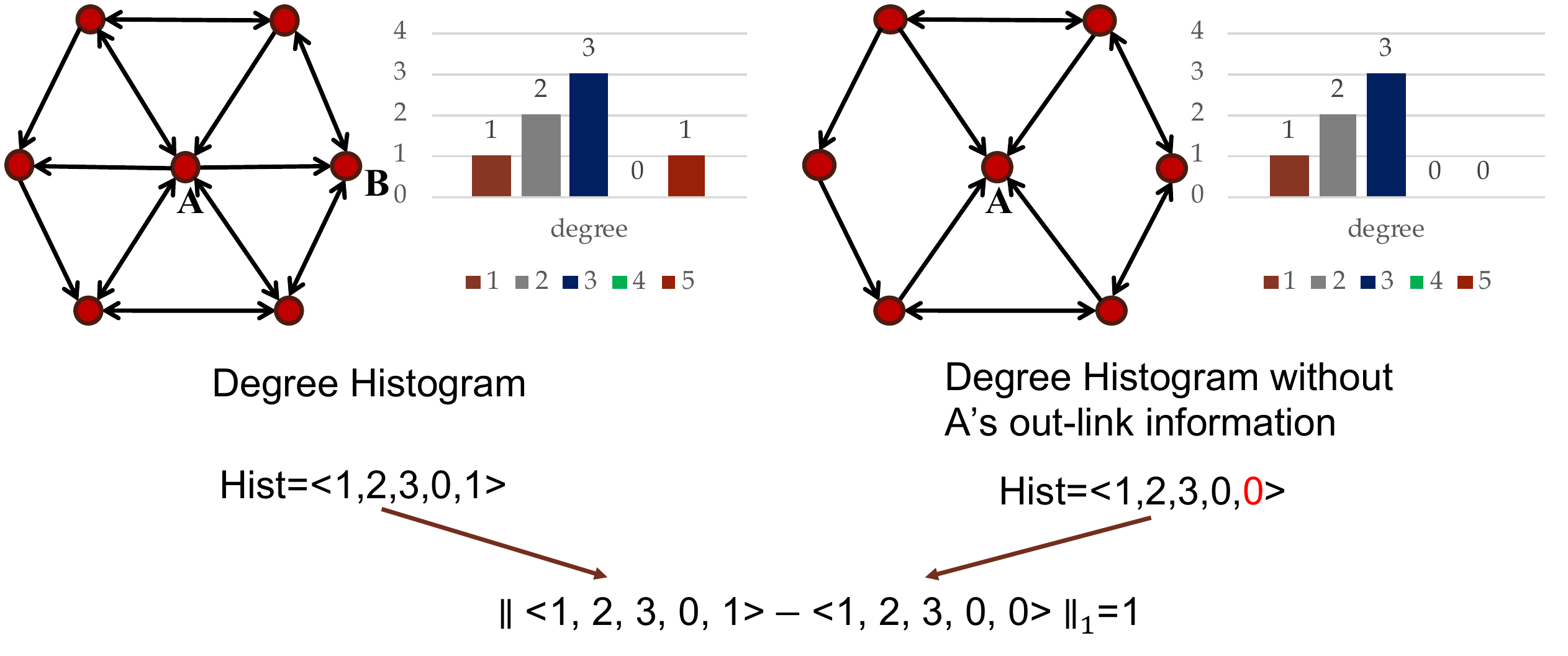}
  \caption{Degree histogram under out-link privacy.}
  \label{fig:11}
\end{figure}

To obtain differentially private results in degree distribution analysis, a number of techniques are proposed, such as \emph{post-processing}~\cite{hrms,HLM}, \emph{projection} (also known as \emph{bounded degree})~\cite{knrs,dll,MPSJ,slmv}, \emph{Lipschitz extension}~\cite{rs}, \emph{Erd\"{o}s-R\'{e}nyi graph}~\cite{SUJ} and  \emph{random matrix projection}~\cite{ALJ}. Post-processing and projection are most commonly used. 

\subsubsection{Post-Processing Techniques} 

Hay~\textit{et~al.}~\cite{hrms} proposed a post-processing technique to boost the accuracy of the existing differentially private algorithms. The key idea is to find a new set of answers that is the ``closest'' to the set of noisy ones returned from differentially private algorithms by means of ``constrained inference'' for better accuracy, that is, enforcing consistency constraints among the noisy query results. 
It involves three steps. First, an analyst sends to the data owner a set of queries with constraints holding among the corresponding answers for a given task. Then the data owner replies to the set of queries using standard differentially private algorithms. In the third step, the analyst post-processes the set of noisy answers with constrained inference to resolve the possible inconsistencies among the noisy answers for the purpose of finding a new set of answers that is the closest to the old one while satisfying the consistency constraints. Here, "closest" is measured in $L_{2}$ distance, and the result is a minimum $L_{2}$ solution. This technique can be viewed as an instance of linear regression. 

Let us use an example to illustrate the procedure. Suppose an analyst needs answers to the total number of students $x_{t}$, the numbers of students $x_{A}$, $x_{B}$, $x_{C}$, $x_{D}$ and $x_{F}$, respectively, receiving grades $A$, $B$, $C$, $D$ and $F$, and the number of passing students $x_{p}$, from a private student database. Intuitively the analyst can obtain differentially private answers to $(x_{A}, x_{B}, x_{C}, x_{D}, x_{F})$, and then use them to compute those for $x_{t}$ and $x_{p}$. Nevertheless, based on the post-processing approach proposed in~\cite{hrms}, the analyst first requests differentially private answers to all queries $x_{t}, x_{p}, x_{A}, x_{B}, x_{C}, x_{D}, x_{F}$, then applies the two constraints $x_{t}=x_{p}+x_{F}$ and $x_{p}=x_{A} + x_{B} + x_{C} + x_{D}$ to derive more accurate answers for $x_t$ and $x_p$. Hay~\textit{et~al.}~\cite{hrms} claimed that the above post-processing technique does not sacrifice privacy. 

Hay~\textit{et~al.}~\cite{HLM}  adapted the definition of differential privacy to graph-structure data and proposed a differentially private 
algorithm based on the post-processing technique proposed in~\cite{hrms} to obtain an approximation of a graph's degree distribution. The authors provided the minimum $L_{2}$ solution to the degree distribution query. The basic idea is to obtain the query results of a graph's degree sequence in a non-decreasing order, then transform them to a degree distribution by counting the frequency of each degree. Let $S$ denote the degree sequence query $S=\langle deg(1), \ldots, deg(n)\rangle$, of which $deg(i)$ denotes the $i^{th}$ smallest degree in $G$. For example, assume that the degrees of a five-node graph are $\{3,3,3,2,1\}$, then $S=\langle1,2,3,3,3\rangle$. Let $\tilde{S}$ denote the sorted results of the differentially private algorithm seeking the degree of each node. Since the degrees are positioned in a sorted order, $S$ is constrained, which can be denoted by $S[i]\le S[i+1]$ for $1\leq i<n$. Then the minimum $L_{2}$ solution $\bar{S}$ is obtained by applying constrained inference to $\tilde{S}$. 

\subsubsection{Bounded Degree Techniques} 

Kasiviswanathan~\textit{et~al.}~\cite{knrs} proposed a carefully-designed projection scheme mapping an input graph to a bounded degree graph to obtain the degree distribution of the original one under node privacy. Aiming at obtaining statistical information with low sensitivity, the original network is projected to a set of graphs whose maximum degree is lower than a certain threshold. In a bounded degree graph, node privacy is easier to achieve as the sensitivity can be much smaller for a given query function. When the degree threshold is carefully chosen for realistic networks, such a transformation leaks little information. Two families of random distributions are adopted for the noise: Laplace distributions with global sensitivity and Cauchy distributions with smooth sensitivity. The key difficulty of this approach lies in that the projection itself may be sensitive to the change caused by a single node in the original graph. Thus, the process of projection should be ``smooth'' enough to ensure the privacy-preservation property of the entire algorithm. Two different techniques were proposed in~\cite{knrs}. The first one defines \emph{tailored projection operators}, which have low sensitivity and protect information for specific statistics. The second one is a ``na\"{i}ve'' projection that just simply discards the high-degree nodes in a graph. Interestingly, the na\"{i}ve projection allows the design of algorithms that can bound the local sensitivity of the projected graph and the development of a generic reduction technique that enables differentially private algorithms for bounded-degree graphs.

Day~\textit{et~al.}~\cite{dll} proposed an edge-addition based graph projection method to reduce the sensitivity of the graph degree distribution problem under node privacy. This improved projection technique preserves more information than the previous ones. It was proved in~\cite{dll} that the degree histogram under the projected graph has sensitivity $2\theta+1$ for a $\theta$-bounded graph in which the maximum degree is $\theta$. Based on this sensitivity bound, two approaches, namely $(\theta, \Omega)$-Histogram and $\theta$-Cumulative Histogram, for degree histograms were proposed under node privacy. Macwan~\textit{et~al.}~\cite{MPSJ} adopted the same method of edge-addition~\cite{dll} to reduce the sensitivity of the node degree histogram. Note that the existing projection-based approaches cannot yield good utility for continual privacy-preserving releases of graph statistics. To tackle this challenge, Song~\textit{et~al.}~\cite{ slmv} proposed a differentially private solution to continually release degree distributions with a consideration on privacy-accuracy tradeoff, assuming that there is an upper bound on the maximum degree of the nodes in the whole graph sequence.

\subsubsection{Other Techniques} 

Raskhodnikova~\textit{et~al.}~\cite{rs} proposed an approximation of the graph degree distribution by making use of the Lipschitz extension and the generalized exponential mechanism under node privacy. Sealfon~\textit{et~al.}~\cite{SUJ} developed a simple, computationally efficient algorithm for estimating the parameter of an Erd\"{o}s-R\'{e}nyi graph under node privacy. This algorithm optimally estimates the edge-density of any graph whose degree distribution is concentrated on a small interval. Ahmed~\textit{et~al.}~\cite{ALJ} presented a random matrix approach to social network data publishing, which achieves differential privacy with storage and computational efficiency by reducing the dimensionality of adjacency matrices with random projection. The key idea is to first randomly project each row of an adjacency matrix into a low-dimensional space, then perturb the projected matrix with random noise, and finally publish the projected and perturbed matrix. The random projection retains the graph matrix's top eigenvectors. As both random projection and random perturbation can preserve differential privacy with a small amount of noise, data utility can be improved.

\subsection{Subgraph Counting}

Given an input graph $G$ and a query graph $H$, a subgraph counting query asks for the number of isomorphic copies of $H$ in $G$. Example subgraphs include triangles, $k$-triangles, $k$-stars, and $k$-cliques, where a $k$-triangle consists of $k$ triangles sharing one common edge, a $k$-star is composed of a central node connecting to $k$ other nodes, and a $k$-clique is a clique with $k$ vertices. Figure~\ref{fig:12} demonstrates these subgraphs. 

Note that subgraph counting counts the copies of a subgraph. Therefore a node of degree $d\ge k$ contributes $\binom{d}{k}$ to $k$-star counting. Figure~\ref{fig:13} presents a few examples of subgraph counting. We consider the counting problems of triangle $k$-star, and $k$-triangle in this section, and denote them respectively by $f_{\triangle}$, $f_{k*}$ and $f_{k\triangle}$. These counting results are keys to many descriptive graph statistics that are used to describe and compare graph properties and structures. For example, the clustering coefficient of a graph is the ratio of $3f_{\triangle}$ over $f_{2*}$. 

\begin{figure}[t]
 \centering
  \includegraphics[width=0.6\textwidth]{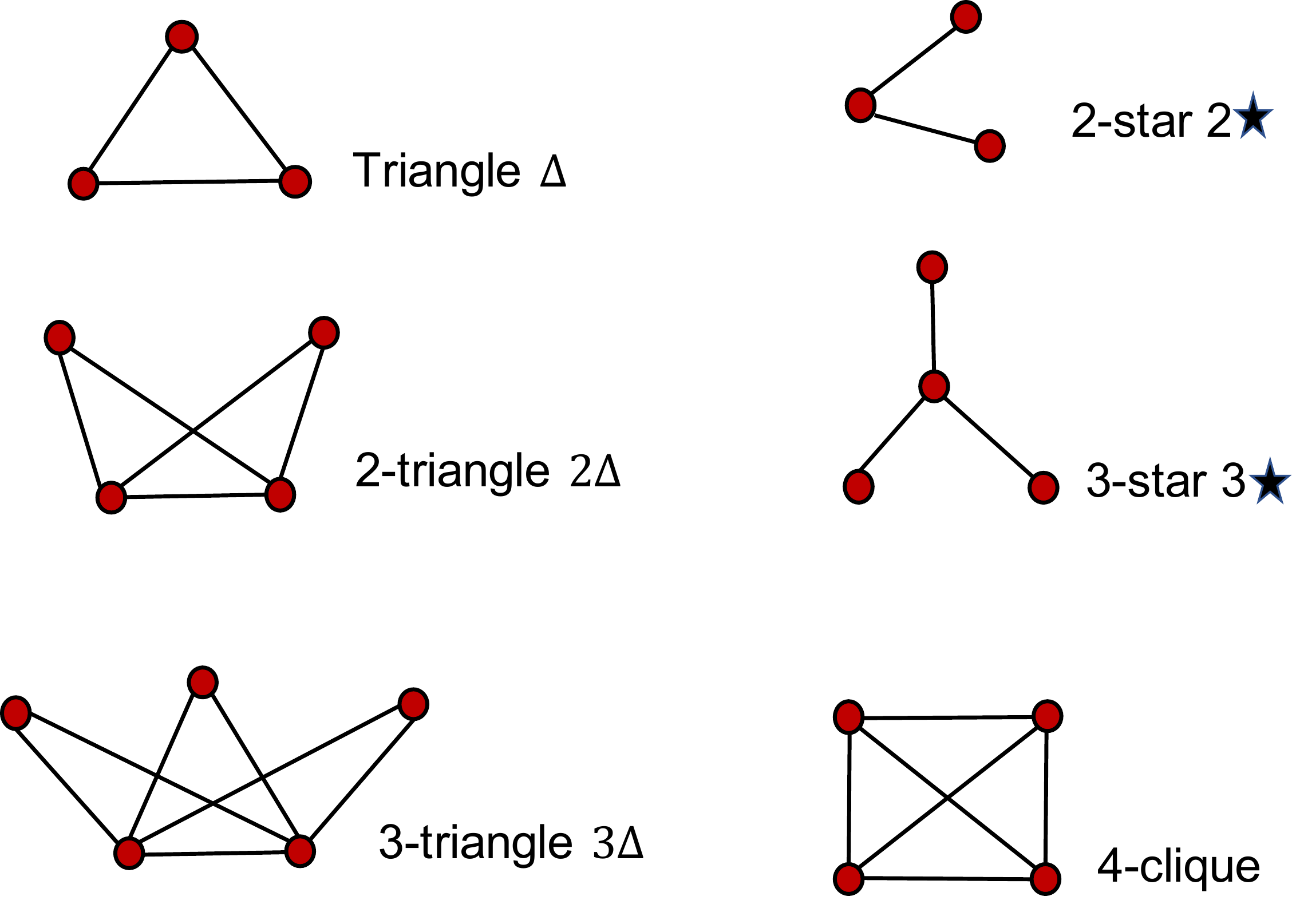}
  \caption{Examples of subgraphs.}
  \label{fig:12}
\end{figure}

\begin{figure}[t]
 \centering
  \includegraphics[width=0.6\textwidth]{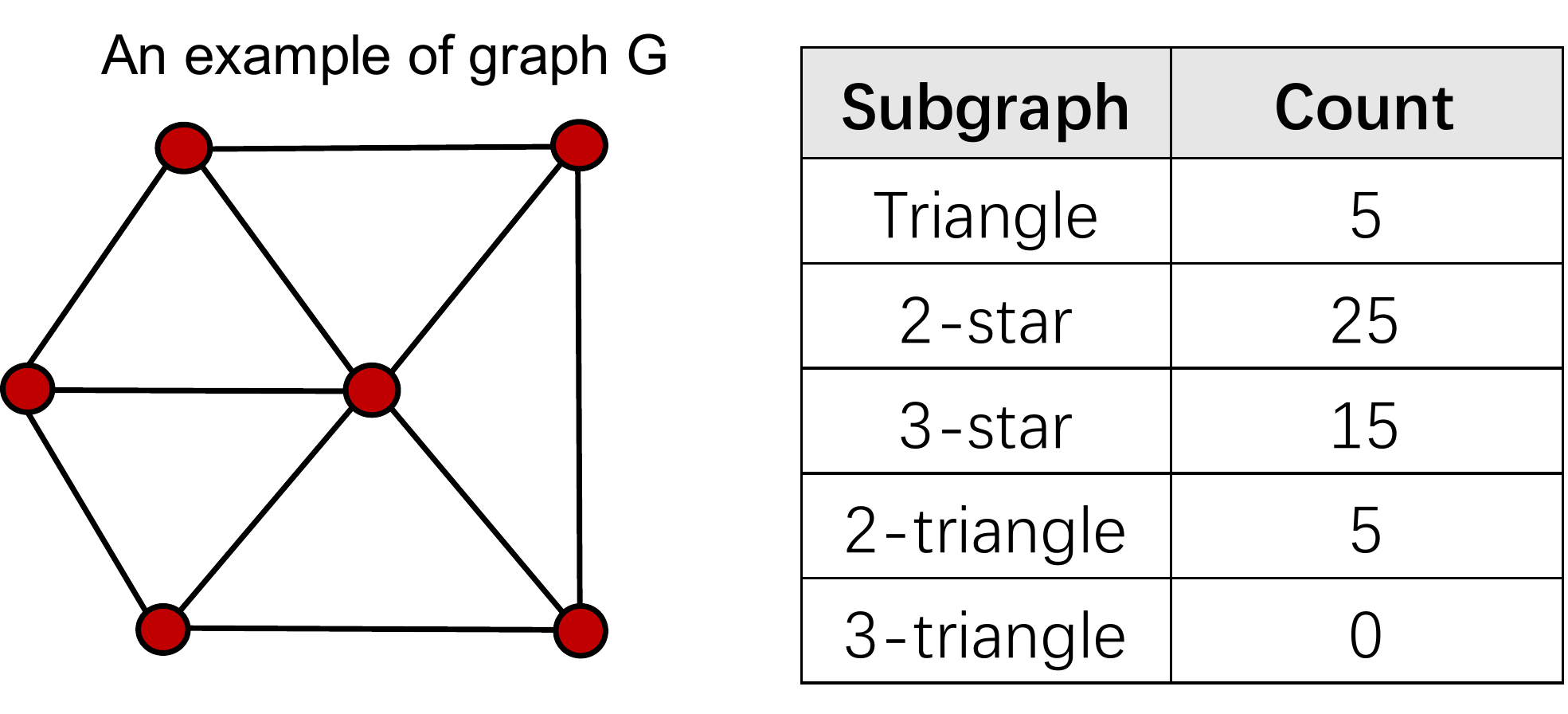}
  \caption{Examples of subgraph countings.}
  \label{fig:13}
\end{figure}

Subgraph counting queries generally have different privacy characteristics and high global sensitivities. To realize differential privacy, it is necessary to add a large amount of noise, which may lead to serious query result distortions. Therefore, a smooth upper bound of the local sensitivity is usually used to determine the noise magnitude. Additionally, truncation, Lipschitz extension and ladder function have been adopted in literature \cite{KRS,knrs,zcp} to achieve differential privacy while improving the counting performance. 

Before summarizing the state-of-the-art techniques, let us introduce some notations.
For an undirected graph with $n$ nodes, the adjacency matrix is $X=(x_{ij})$, where $x_{ii}=0$ for all $i\in[n]$. Let $a_{ij}$ denote the number of common neighbors shared by a particular pair of vertices $i$ and $j$, that is, $a_{ij}=\sum_{l\in [n]}x_{il}\cdot x_{lj}$. Let $b_{ij}$ denote the number of vertices connected only to one of the two vertices $i$ and $j$, that is, $b_{ij}=\sum_{l\in [n]}x_{il}\bigoplus x_{lj}$. Denote by $d(G,G')$ the distance between two $n$-vertex graphs $G$ and $G'$, which is the number of edges they differ. Graph $G$ and $G'$ are neighbors if $d(G,G')=1$. Let $LS_{\triangle}$, $LS_{k\triangle}$ and $LS_{k*}$ denote the local sensitivities of $f_{\triangle}$, $f_{k\triangle}$ and $f_{k*}$, respectively. Denote by $S^{*}_{\triangle,\beta}$, $S^{*}_{k\triangle,\beta}$ and $S^{*}_{k*,\beta}$ the smooth sensitivities of $f_{\triangle}$, $f_{k\triangle}$ and $f_{k*}$, respectively.

\subsubsection{Triangle Counting}


As mentioned earlier, node privacy is a strong privacy guarantee, so it is not feasible to obtain a triangle counting satisfying node privacy in most cases. At the worst case, adding a vertex to a complete $n$-node graph brings $\binom{n}{2}$ new triangles. Since this change depends on the size of the graph, the global sensitivity of triangle counting is unbounded. 
Moreover, triangle counting is also not feasible under edge privacy as in the worst case, deleting one edge from an $n$-node graph deletes $n-2$ triangles. 
Although the global sensitivity of a triangle counting query is not bounded, its local sensitivity for some specific graphs is bounded under edge privacy. Thus, smooth sensitivity~\cite{NRS,KRS} can be adopted to achieve differential privacy. 
In the following, we briefly summarize edge and node differentially private algorithms as well as other techniques to achieve differential privacy in triangle counting.

\paragraph{Edge Differentially Private Algorithms} 

Nissim~\textit{et~al.}~\cite{NRS} introduced an approach to calculate the smooth sensitivity of triangle counting and provided the cost of a minimum spanning tree under edge privacy. The local sensitivity of $f_{\triangle}$ is $LS_{f_{\triangle}}=\max_{i,j\in [n]}a_{ij}$, the global sensitivity is $GS_{f_{\triangle}}=n-2$, while $LS_{f_{\triangle}}$ at distance $s$ is $LS_{f_{\triangle}}^{(s)}=\max\limits_{i\neq j;i,j\in n}c_{ij}(s)$, where $c_{ij}(s)=\min(a_{ij}+\lfloor \frac{s+\min(s, b_{ij})}{2}\rfloor,n-2)$. The $\beta$-smooth sensitivity of $f_{\triangle}$ has time complexity $O(M(n))$, where $M(n)$ is the time required for multiplying two matrices of size $n\times n$.

Karwa~\textit{et~al.}~\cite{KRS} presented an efficient algorithm for outputting approximate answers to subgraph counting queries, such as triangle counting, $k$-star counting and $k$-triangle counting. These algorithms satisfy edge privacy and can be regarded as an extension of the algorithm in~\cite{NRS} to a bigger class of subgraph counting problems with privacy guarantees and better accuracy.

Sala~\textit{et~al.}~\cite{SZW} proposed a differentially private graph model called \emph{Pygmalion} to generate synthetic graphs. They adopted the $dK$-graph model and its statistical series as the query function. The $dK$-graph model extracts the detailed structure of a graph into degree correlation statistics, and outputs a synthetic graph using the $dK$-series values. A $dK$-series is the degree distribution of connected components of certain size within a target graph. Here, the $dK$-series is a graph transformation function. Sala~\textit{et~al.}~\cite{SZW} first proved that the $dK$-series has a high sensitivity, then proposed a partitioning approach to group tuples with similar degrees, which effectively reduces the noise magnitude and achieves a desired privacy guarantee.

Zhang~\textit{et~al.}~\cite{zcp} proposed an approach of specifying a probability distribution over possible outputs to maximize the utility of an input graph while providing a privacy guarantee. They applied a ladder function to the subgraph counting problems of triangle, $k$-star and $k$-clique, and achieved high accuracy with efficient time complexities. 

Gupta~\textit{et~al.}~\cite{gru} considered the problem of approximately publishing the cut function of a graph under edge privacy. They proposed a generic framework of converting iterative database construction algorithms into privatized query publishing approaches under non-interactive and interactive settings.

Qin~\textit{et~al.}~\cite{ZTY} made an effort to ensure individual's local differential privacy while gathering structural information to generate synthetic social graphs. They proposed a multi-phase technique of LDPGen, which incrementally clusters structurally similar users via refining parameters into different partitions. Specifically, whenever a user reports information, LDPGen deliberately injects noise to guarantee local differential privacy. Moreover, LDPGen derives optimal parameters to cluster structurally similar users together. After obtaining a good clustering, LDPGen constructs a synthetic social network by adopting the existing Chung-Lu social graph generation model~\cite{aie2000}.

\paragraph{Node Differentially Private Algorithms} 

Since node privacy is a strong privacy guarantee, a large amount of noise needs to be added, leading to a dramatic distortion of the graph structure and a poor utility. One of the most widely adopted mechanisms is the generic reduction to privacy over a bounded-degree graph. If a graph is known to have a maximum degree of $d$, deleting or adding a node may affect $\binom{d}{2}$ triangles at most. For graphs whose maximum degree is greater than $d$, high-degree nodes can be deleted to get a graph with a maximum degree falling within a threshold. The number of triangles of this bounded-degree graph can be a good approximation to the true query answer. Therefore, networks with a small number of large degree nodes can adopt this approach to achieve node privacy for triangle counting.


Kasiviswanathan~\textit{et~al.}~\cite{knrs} proposed algorithms for releasing statistics of graph data under node privacy. On the basis of smooth sensitivity of truncation, they presented a generic reduction mechanism in order to apply differentially private algorithms for bounded-degree graphs to arbitrary graphs, that is, just simply removing the nodes with high degrees. A continual privacy-preserving release of subgraph counting under node privacy was investigated in~\cite{slmv}, which assumes that there is a publicly known upper bound on the maximum degree of the nodes in the graphs. 

Jeremiah~\textit{et~al.}~\cite{BBD} proposed the definition of \emph{restricted sensitivity}, which can improve the accuracy of differential privacy compared with global sensitivity and smooth sensitivity. Two important query classes, namely subgraph counting and local profile matching of social networks, were analyzed. It was proved that the restricted sensitivities of these two kinds of queries are much lower than those under smooth sensitivity. More importantly, when computing the smooth sensitivity involves higher computational complexity and lower efficiency, restricted sensitivity performs better. 

\paragraph{Other Types of Privacy}

Rastogi~\textit{et~al.}~\cite{rhms} considered general privacy-preserving social network queries including subgraph counting. They proposed a relaxation of edge privacy, called a \emph{theoretic standard of adversarial privacy}. Their algorithm can release more general graph statistics than the algorithms in~\cite{NRS}, which only deal with triangles. However, the assumption on adversarial privacy puts some limits on the applicability of this privacy definition~\cite{rhms}. 

Task~\textit{et~al.}~\cite{TCS} proposed two differential privacy standards, i.e., \emph{out-link privacy} and \emph{partition privacy}, over network data. They also introduced two algorithms respectively satisfying the two privacy standards to release approximate results of degree distribution query, triangle counting and centrality counting. It was demonstrated that partition privacy can provide stronger privacy guarantee with less noise when cross-analyzing multiple social networks. 

Gehrke~\textit{et~al.}~\cite{glp} presented a zero-knowledge based privacy definition, which is stronger than differential privacy. They constructed a zero-knowledge private mechanism to release the social graph structure information such as the average degree and the distance to connectivity.

Sun~\textit{et~al.}~\cite{SXK} pointed out that it is insufficient to apply local differential privacy to protect all network participants when collecting extended local views (ELV). The main problem lies in that each individual has its own local privacy budget, which covers its own ELV regardless of those the neighbors in her ELV have. To prevent this attack, a novel decentralized differential privacy (DDP) mechanism was proposed, which demands each participant to consider not only its own privacy, but also those of the neighbors in its ELV. Towards this goal, a multi-phase mechanism under DDP was developed, which allows an analyst to better estimate subgraph counting. In this framework, an analyst first queries each individual's minimum noise scale, which must be performed under DDP since it relies on the local graph structure and is private. Then, the analyst calculates the minimum noise scale for the whole network and gathers subgraph counting accordingly.

\subsubsection{\textbf{$k$-Star Counting}}


Karwa~\textit{et~al.}~\cite{KRS} extended the approach in~\cite{NRS} to the $k$-star counting query and proposed how to compute the local sensitivity and smooth sensitivity of $f_{k*}$. They proved that these two sensitivity values of $k$-star counting are equal, that is, $S^*_{k*, \beta}(G)=LS_{k*}(G)$ when $d_{\max} \geq \max\{ k, (k-1)(\frac{1-\beta}{\beta}) \}$, of which $d_{\max}$  is the largest degree in $G$.

\nop{
Let $f_{k*}(G)$ be the number of $k$-stars in $G$, then $f_{k*}(G)=\sum_{i\in[n]}\binom{d_{i}}{k}$, where $d_{i}$ is the degree of node $i$. For all $k\geq 2$, the local sensitivity of $f_{k*}(G)$ is $LS_{k*}(G)=\max\limits_{i\neq j,i,j\in [n]}(\binom{d_{i}-x_{ij}}{k-1}+\binom{d_{j}-x_{ij}}{k-1})$. Let $C(a)$ denote $\binom{a}{k-1}$, $d'_{i}=d_{i}-x_{ij}$, and $B_{i}=n-2-d'_{i}$; similarly $d_{j'}=d_{j}-x_{ji}$ and $B_{j}=n-2-d'_{j}$. 
We have $LS^{(t)}_{k*}(G)=\max\limits_{(i,j):d_{i}\geq d_{j}}LS_{ij}^{(t)}(G)$, where $LS_{ij}^{(t)}(G)$ is the local sensitivity of  $f_{k*}$ at distance $t$ over an edge $(i,j)$. If $d_{i}\ge d_{j}$, we have 
%
\begin{eqnarray}
LS_{ij}^{(t)}(G)= \nonumber\\
\begin{cases}
C(d'_{i}+t)+C(d'_{j}) & \text{if}\quad t \leq B_{i} \\ 
 C(n-2)+C(d'_{i}+t-B_{j})& \text{if} \quad t \in (B_{i}, B_{i}+B_{j}) \\ 
2C(n-2)& \text{if} \quad t\ge B_{i}+B_{j}
\end{cases}
\end{eqnarray}
}
%

Kasiviswanathan~\textit{et~al.}~\cite{knrs} proposed an $(\epsilon,\delta)$-node differentially private algorithm with a linear programming (LP) based function for the special case of the subgraph $H$ having $3$ nodes, e.g., $H$ can be a triangle or a $2$-star. If $f_{H}(G)$ (the number of copies of $H$ in $G$) is relatively large, the Laplace mechanism provides an accurate estimate. The release of $f_{H}(G)$ is more accurate with the LP-based function when $f_{H}(G)$ is smaller. 

Zhang~\textit{et~al.}~\cite{zcp} presented a ladder function and applied it to the $k$-star query under edge privacy. The ladder function relies on a carefully designed probability distribution that can maximize the probability of outputting true answers and minimize that of outputting the answers that are far from the true answers. In addition, to achieve differential privacy, it is constrained that the probabilities of outputting a value for the input graph $g$ and its neighbor $g'$ should be very close. The authors adopted the concept of ``local sensitivity at distance $t$'' in~\cite{knrs} to create a ladder function. In fact, the upper bound of the ``local sensitivity at distance $t$'' was used as the ladder function for the $f_{k*}$ query. 

\nop{
The global sensitivity of $f_{k*}$ is $GS=2\binom{n-2}{k-1}$ while the local sensitivity at distance $t$ of $f_{k*}$ is $LS(g,t)=\max_{i,j}LS_{ij}(g,t)$, where
%
\begin{equation}
LS_{ij}(g,t)=
\begin{cases}
\vspace{2mm}
\binom{\bar{d_{i}}+t}{k-1}+\binom{\bar{d_{j}}}{k-1} & \text{if}\quad t \leq B_{i} \\ 
\vspace{2mm}
 \binom{n-2}{k-1}+\binom{\bar{d_{i}}+t-B_{i}}{k-1}& \text{if} \quad t \in (B_{i}, B_{i}+B_{j}] \\ 
 \vspace{2mm}
2\binom{n-2}{k-1}& \text{if} \quad t> B_{i}+B_{j}
\end{cases}
\end{equation}
in which $\bar{d_{i}}=d_{i}-x_{ij}$ and $B_{i}=n-2-\bar{d_{i}}$; and $\bar{d_{j}}=d_{j}-x_{ji}$ and $B_{j}=n-2-\bar{d_{j}}$.
}

\subsubsection{\textbf{$k$-Triangle Counting}}


When triangle counting is extended to $k$-triangle counting, the problem becomes complicated as it is NP-hard to calculate the smooth sensitivity of $k$-triangle counting. Therefore, existing approaches mainly focus on a small $k$, while the counting query of $f_{k\triangle}$ itself is hard.

An approach was proposed in~\cite{KRS}, whose main idea is to compute $(\epsilon, \delta)$-differential privacy (edge privacy) by adding noise proportional to a second-order local sensitivity instead of a ``smooth'' upper bound. Since $LS_{k\triangle}$ has a high sensitivity, it cannot be directly adopted with the Laplace mechanism. Therefore, $LS'$, the local sensitivity of $LS_{k\triangle}$, was employed. It was demonstrated that $LS'$ is a deterministic function of a quantity with global sensitivity $1$, based on which the query results can be published with less noise. 
Another approach was presented by Zhang~\textit{et~al.}~\cite{zcp}, which provided a ladder function for $k$-triangle counting under edge privacy. 

\nop{
Let $N_{ij}$ denote the common neighbor set of the nodes $i$ and $j$, that is $N_{ij}=\{l\in [n]\mid x_{il}\cdot x_{lj}=1\}$; correspondingly, we have $|N_{ij}|=a_{ij}$. The local sensitivity of $f_{k\triangle}$ is 
%
\begin {equation}
LS_{k\triangle}(G)=\max_{i,j\in [n],i\neq j}LS_{ij}(G)
\label{equ:ls}
\end{equation}
where
\begin {equation}
LS_{ij}(G)=\binom{a_{ij}}{k}+\sum\limits_{l\in N_{ij}}\binom{a_{il}-x_{ij}}{k-1}+\binom{a_{lj}-x_{ij}}{k-1}
\label{equ:ls'}
\end{equation}
Here $LS'$ is the local sensitivity of the local sensitivity function $LS_{k\triangle}(G)$. Let $a_{\max}=\max\limits_{i,j\in n,i\neq j}a_{ij}$, then we have
\begin {equation}\nonumber
LS'(G)\leq 3\binom{a_{\max}}{k-1}+a_{\max}\binom{a_{\max}}{k-2}
\end{equation}
By computing the added noise magnitude according to $LS'$, one can get $f_{k\triangle}$.

As mentioned in the section of $k$-star counting, Zhang~\textit{et~al.} also provided a ladder function for $k$-triangle counting in~\cite{zcp} under edge privacy. 
}

\nop{
The global sensitivity of $f_{k_{k\triangle}}$ is $GS=\binom{n-2}{k}+2(n-2)\binom{n-3}{k-1}$. Based on \eqref{equ:ls} and \eqref{equ:ls'}, one can design a ladder function for $k$-triangle counting, which is shown below:
\begin{equation}
I_{t}(G)=\min(LS_{k\triangle}(G)+\sum_{i=0}^{t-1}U(a_{m}+i),GS)
\end{equation}
where $U(a)=3\binom{a}{k-1}+a\binom{a}{k-2}$, and $a_{m}$ is the maximum number of common neighbors in graph $G$.}

\subsection{Edge Weights}
 
In social networks, social relations are modeled on edges with weights. An edge may reveal different sensitive information between individuals, such as the communication cost, the interaction frequency between two social network users, the price of a commercial trade or the similarity between two organizations. Thus, releasing edge weights must be done in a privacy preserving manner. Table~\ref{tab:5} summarizes the most popular exiting differentially private edge weight algorithms in social networks. 

Liu~\textit{et~al.}~\cite{lwl} studied the problems of protecting privacy in edge weights and preserving the utility of statistics of shortest paths between nodes. They proposed two edge privacy-preserving approaches, namely \emph{greedy perturbation} and \emph{Gaussian randomization multiplication}. The former mainly focuses on preserving the length of the perturbed shortest paths and the latter retains the same shortest paths before and after perturbation. 

Das~\textit{et~al.}~\cite{dee} conducted edge weight anonymization in social graphs. They developed a linear programming model to protect graph characteristics such as shortest paths, minimum spanning trees and $k$-nearest neighbors, which can be formalized as linear functions of the edge weights. 

Costea~\textit{et~al.}~\cite{CBR} considered differential privacy protection to the edge weights assuming that the graph structure is public and available to users without modification while the edge weights are private. They employed the Dijkstra algorithm to get the shortest paths for protection quality evaluation. 

Last, Li~\textit{et~al.}~\cite{LYS} treated the edge-weight sequence as an unattributed histogram by merging all barrels with the same count into one group and thus ensured $k$-indistinguishability among groups. They proposed an approach with Laplace noise added to every edge weight to improve accuracy and utility of the published data.

\begin{table}[t]
\arrayrulewidth=0.5pt
\tabcolsep=14pt  
\caption{Summary on Existing Edge Weight Preservation Techniques}
\centering
\begin{tabular}{ll}
\hline
     & \textbf{Approach}                                                        \\
      \hline
         \hline
\begin{tabular}[c]{@{}l@{}}~\cite{lwl}\end{tabular}               & \begin{tabular}[c]{@{}l@{}}Gaussian randomization multiplication; \\Greedy perturbation\end{tabular}                     \\ \hline
\begin{tabular}[c]{@{}l@{}}\cite{dee}\end{tabular} & \begin{tabular}[c]{@{}l@{}} Linear programming model\end{tabular}                   \\ \hline
\begin{tabular}[c]{@{}l@{}}\cite{CBR}\end{tabular}  & \begin{tabular}[c]{@{}l@{}} Edge weight-count; Laplace perturbation\end{tabular} \\ \hline
\begin{tabular}[c]{@{}l@{}}\cite{LYS}\end{tabular}  & \begin{tabular}[c]{@{}l@{}} Edge weight-unattributed histogram; \\$k$-indistinguishability\end{tabular}      \\ \hline
\end{tabular}
\label{tab:5}
\end{table}

\subsection{Summary}

Social networks contain information about social users, their attributes as well as social relationships, which are usually deemed sensitive. The release of such information may bring significant privacy concerns or even damages to personal reputation and properties if the protection on sensitive information is not sufficiently strong. In this section, we discussed degree distribution, subgraph counting (triangle, $k$-star and $k$-triangle) and edge weights, the most popular graph analysis techniques in social networks. Note that there exist other statistics on graph structures but the basic methods and ideas are similar and thus are omitted here.  

According to our analysis, most existing differentially private algorithms cannot obtain good utility (high accuracy) for large-scale and complex graph structures. Moreover, the complexity of the differentially private algorithms are generally high or even NP-hard due to the complexity of computing (smooth) sensitivities. In some cases such as $k$-triangle counting, even the structure query itself is already NP-hard. 

To the best of our knowledge, no high sensitivity problem was reported in local differential privacy. However, it is a great challenge for data collectors to reconstruct a graph structure with high utility based on the disturbed or local graph data, that is, to ensure the correlations between the original data when the perturbation process of each user is independent of each other. Furthermore, if we only collect graph statistics, such as node degrees and subgraph counting, to generate composite graphs, the output graph may not retain the important characteristics of the original one and thus may reduce graph utility.

\section{Conclusions and Future Directions}\label{sec:saf}

In this article, we provide a survey on differential privacy foundations and applications in protecting the privacy of social network analytical results. We explain the underlying design principles of different mechanisms and present the state-of-the-art research results. To achieve differential privacy, one needs to specify a privacy budget and calculate the amount of noise to be added to the query results. The privacy budget determines the level of privacy preservation: the smaller, the better the protection. At the same time, the noise magnitude affects the accuracy (utility) of the query results, which should be minimized provided that sufficient privacy protection is achieved. Noise magnitude is derived from sensitivity and privacy budget. When global sensitivity is high, smooth sensitivity may be employed instead. 

The research on differential privacy is developing fast, and its applications in social network analysis enjoy stronger and stronger interest from industry and academia. In the following we discuss a few open research problems in differential privacy technologies for social network analysis.

\subsection{Differential Privacy for Complex and Correlated Social Network Data}
In social networks a user often has relationships with many others at different levels. Thus, network structures are often complex. Since query sensitivities in social networks are usually high, much noise has to be added to query results to achieve differential privacy.  Nevertheless, the noise may significantly affect the output data utility. In addition, it may be hard to effectively compute sensitivities, either global or smooth, precise or approximate, as the computational complexity may be too high (or even NP-hard) to be practical for many complex social network analysis queries. 
Even though a large number of studies reviewed earlier focus on how to apply differential privacy to complex social structure queries, most of them are limited to ``small'' queries, such as a small $k$ in $k$-star and $k$-triangle counting. It remains a great challenge to employ traditional differential privacy for complex graph queries. 

Moreover, in social networks, social correlations are usually strong as behaviors and attributes of adjacent nodes are often strongly related. For example, adjacent users may have the same attributes with a high probability. Therefore, the private attributes of a social network node may be inferred by exploring the publicized attributes of its neighbors which share common interests~\cite{CAK2012}. The social relations, that is, the edges in a social network, are often not independent, as the social relationship between two nodes may depend on a third node that is a common neighbor. 
To address dependency in data, dependent differential privacy has attracted a lot of attention in recent years~\cite{LCM,ZZP}. Nevertheless, applying dependent differential privacy to social networks remains to be a grand open challenge due to high dependencies and complex social structures.


To tackle the challenges, one possible direction is the transformation techniques. 
For example, we may consider adding a sampling process to transform an original graph data to one in a different domain such that the data tuples become independent and sparse and thus traditional differential privacy can be applied. This is motivated by the random but uniform sampling step in~\cite{LQS}. The non-uniform compressive sampling technique~\cite{CandesCompressiveSampling,SparseTargetCounting-Infocom2011,TangTWC13} may be employed as it can realize the required transformation with controlled distortions. 

\subsection{Tradeoff between Privacy Budget and Data Utility}

How to allocate an appropriate privacy budget to achieve sufficient privacy protection on sensitive data and, at the same time, maximize data utility remains a fundamental challenge~\cite{ZhangPUC19}. 
Recently various schemes were developed to investigate the privacy-utility tradeoff based on techniques such as game theory and linear programming~\cite{JMA,cui2019,CRJ,CDNK}.
 
Dwork~\textit{et~al.}~\cite{CDNK} stated that there is little understanding on the optimal value of privacy budget for a practical scenario. Importantly, their interview results obtained from surveying different practitioners regarding how organizations made key choices when implementing differential privacy in practice indicated that there was no clear consensus on how to choose privacy budget, nor agreement on how to approach to the problem. One challenge is to quantify the tradeoff between privacy budget and data utility. 

\subsection{Differentially Private Publishing of High Dimensional Social Network Data}

The unprecedented growth and popularity of online social networks have generated massive high-dimensional data, such as social users' attribute information, healthcare data, location information, trajectory data, and commercial electronic data, which is often published or made available to third parties for analysis, recommendations, targeted advertising and reliable prediction. However, publishing such attribute data may disclose private and sensitive information and result in increasing concerns on privacy violations. Differentially private publishing of such data has received broad attentions. Nevertheless, most differentially private data publishing techniques cannot work effectively for high dimensional data. On the one hand, since the sensitivities of different dimensions vary, evenly distributing the total privacy budget to each dimension degrades the performance. Moreover, the ``Curse of Dimensionality'' leads to two critical problems. First, a dataset containing many dimensions and large attribute domains has a low ``Signal-to-Noise Ratio''~\cite{ZCPC}. 
Second, complex correlations exist between attribute dimensions, making it impossible to directly and independently protect each dimension's privacy. To address these challenges, one may conduct data dimensionality reduction. However, it is hard to maintain the characteristics of high dimensional data to the maximum extent and to prevent private information from being defected during the process of dimensionality reduction.

To address these challenges, Bayesian networks~\cite{ZCPC}, random projection~\cite{XRZ} and various sampling techniques are used to support differentially private high dimensional data publishing~\cite{LXJ,CXZ}. Nevertheless, most of these approaches still cannot work effectively for releasing high-dimensional data in practice as they generally ignore the different roles a dimension may play for a specific query -- one dimension may be more important than another for a particular query. Additionally, one dimension may release more information than another if the same amount of noise is added.  Therefore, how to allocate the total privacy budget to dimensions and optimize privacy protection is query-dependent and should be carefully investigated. Moreover, the underlying distribution of the data may be unknown and the high dimensionality and large attribute domains may skew the distributions of different dimensions, leading to significant perturbations on the published data and thus affecting data utility. Last, dimensionality reduction and noise addition both introduce defection to the published data. How they jointly affect data utility is a tough and open problem.

\subsection{Differentially Private Publishing of Dynamic Data}

Most of the existing differential privacy research focuses on static data publishing. In practice, many datasets, such as online retail data, recommendation system information and trajectory data, are dynamically updated. Representing dynamic social network data as a static graph and discarding temporal information may result in the loss of evolutionary behaviors of social groups. Thus, how to achieve differential private danymic social network data publishing is an important research direction. 

Differential private publishing of dynamic social network data faces two critical challenges:  allocating privacy budget to each data element at each version and handling noise accumulation over continuous data publishing. In an algorithm with multiple sequential queries, the privacy budget may be exhausted after a while based on the notion of composite differential privacy, and thus the promised privacy protection may not be maintained. Therefore, we need a budget allocation strategy that can make the life cycle of privacy budget as long as possible while providing sufficient protection in a composite query. 
Moreover, since each updated data publishing must consider the added noise in the previous one to counter the correlation between the two releases, the cumulative noise increases rapidly as the number of releases increases, resulting in the fast decreasing utility in the published data over time. 


There exist initial efforts on this direction. For example, Chan~\textit{et~al.}~\cite{CSS} and Chen~\textit{et~al.}~\cite{CAC} tackled the continual counting problem and the differential private publishing of sequential data.  However, the proposed approaches do not address the failures caused by early exhaustion of privacy budget. The continual release of degree distributions in degree-bounded graphs was considered in~\cite{ slmv} but the proposed technique yields poor utility.

Generally speaking,
most approaches for differentially private data publishing in a static environment cannot be directly applied to publishing of dynamic data. Although the data released at a certain moment satisfies differential privacy, as the number of updates increases, potentially unbounded, the amount of noise required for each release becomes bigger and bigger, resulting in a large cumulative error and low data utility. Once the privacy budget is exhausted, differential privacy protection cannot be guaranteed. New strategies and mechanisms are highly needed.
\bibliographystyle{abbrv}
\bibliography{ref} 

\end{sloppy}
\end{document}